\documentclass[10pt,a4paper,twocolumn]{IEEEtran}
\def\psscale{0.7}

\usepackage{amssymb,amsmath,amsthm}
\usepackage{setspace}
\usepackage{pstricks}
\usepackage{psfrag}
\usepackage{color}
\usepackage{cite}
\usepackage{graphicx}
\usepackage{wrapfig}
\usepackage{subcaption}
\usepackage{algorithmic}
\usepackage{algorithm}

\def\bsepsilon{{\boldsymbol{\epsilon}}}

\def\bsPi{{\boldsymbol{\Pi}}}
\def\bsPhi{{\boldsymbol{\Phi}}}

\def\bsDelta{{\boldsymbol{\Delta}}}

\def\bsSigma{{\boldsymbol{\Sigma}}}
\def\bsPsi{{\boldsymbol{\Psi}}}

\def\bse{{\boldsymbol{e}}}

\def\bsh{{\boldsymbol{h}}}

\def\bsn{{\boldsymbol{n}}}

\def\bss{{\boldsymbol{s}}}

\def\bsx{{\boldsymbol{x}}}
\def\bsy{{\boldsymbol{y}}}

\def\bsA{{\boldsymbol{A}}}

\def\bsD{{\boldsymbol{D}}}

\def\bsH{{\boldsymbol{H}}}
\def\bsI{{\boldsymbol{I}}}

\def\bsL{{\boldsymbol{L}}}

\def\bsP{{\boldsymbol{P}}}
\def\bsQ{{\boldsymbol{Q}}}
\def\bsR{{\boldsymbol{R}}}

\def\bsU{{\boldsymbol{U}}}
\def\bsV{{\boldsymbol{V}}}

\def\bszero{{\boldsymbol{0}}}

\def\calC{{\mathcal{C}}}

\def\calN{{\mathcal{N}}}

\def\calS{{\mathcal{S}}}

\def\bbR{{\mathbb{R}}}

\def\Nr{{N_\textup{R}}}
\def\Nt{{N_\textup{T}}}

\def\No{{N_\textup{0}}}
\def\Eb{{E_\textup{b}}}
\def\ns{{n_\textup{s}}}

\def\wbP{{\;\overline{\text{\makebox[2mm]{$\bsP\,$}}}\,}}
\def\wtPhi{{\tilde{\bsPhi}}}

\def\wtPsi{{\tilde{\bsPsi}}}
\def\wbPsi{{\bar{\bsPsi}}}
\def\wtH{{\;\text{$\widetilde{\makebox[1ex]{$\bsH$\,}}\,\,$}}}
\def\wbH{{\;\text{$\overline{\makebox[1ex]{$\bsH$}\,}\,$}}}

\newsavebox{\customizedforwidehat}
\savebox{\customizedforwidehat}[1ex]{\raisebox{0mm}[6pt][0mm]{$\bsH$}}
\def\whH{{\;\widehat{\usebox{\customizedforwidehat}\,}\,}}

\newsavebox{\customizedforscriptwidehat}
\savebox{\customizedforscriptwidehat}[1ex]{\raisebox{0mm}[4pt][0mm]{\scriptsize$\bsH$}}
\def\scrwhH{{\;\widehat{\usebox{\customizedforscriptwidehat}\,}\,}}

\def\wtH{{\;\widetilde{\raisebox{0mm}[7pt][0mm]{\makebox[1ex]{$\bsH$\,}}}\;}}
\def\wbH{{\;\overline{\raisebox{0mm}[8pt][0mm]{\makebox[1ex]{$\bsH$\,}}}\;}}

\savebox{\customizedforwidehat}[1ex]{\raisebox{0mm}[7pt][0mm]{$\bsH$}}
\def\whH{{\;\widehat{\usebox{\customizedforwidehat}\,}\,}}
\savebox{\customizedforscriptwidehat}[1ex]{\raisebox{0mm}[5pt][0mm]{\scriptsize$\bsH$}}
\def\scrwhH{{\;\widehat{\usebox{\customizedforscriptwidehat}\,}\,}}

\newcommand{\norm}[1]{\left\lVert{#1}\right\rVert}
\newcommand{\inv}[2][0mm]{{#2}^{\text{\raisebox{#1}{\makebox[2mm]{\tiny$-1$}}}}}

\DeclareMathOperator*{\argmin}{argmin}
\DeclareMathOperator{\diag}{diag}
\DeclareMathAlphabet{\CMmathcal}{OMS}{cmsy}{m}{n}
\renewcommand{\mathcal}[1]{\CMmathcal{#1}}

\renewcommand{\exp}[1]{\mathrm{exp}\left(#1\right)}
\newcommand{\expp}[0]{\mathrm{exp}}
\newcommand{\figref}[1]{Fig.~\ref{#1}}

\newcommand{\secref}[1]{Sec.~\ref{#1}}
\newcommand{\algref}[1]{Alg.~\ref{#1}}
\newcommand{\appref}[1]{App.~\ref{#1}}
\newcommand{\tabref}[1]{Tab.~\ref{#1}}
\renewcommand{\Re}[1]{\mathrm{Re}\left\{#1\right\}}
\renewcommand{\Im}[1]{\mathrm{Im}\left\{#1\right\}}
\newcommand{\Var}[2][]{\mathrm{Var}_{#1}\left\{#2\right\}}
\newcommand{\Exp}[2][]{\mathbb{E}_{#1}\!\!\left\{#2\right\}}

\author{ \IEEEauthorblockN{Mirsad \v{C}irki\'{c} and Erik G. Larsson}

  \IEEEauthorblockA{
    Communication Systems Division\\%
    Department of Electrical Engineering (ISY)\\%
    Link\"oping University, SE-581 83 Link\"oping, Sweden\\%
    \texttt{\{mirsad.cirkic,erik.g.larsson\}@liu.se}}
  
  \thanks{This work was supported in part by the Swedish Research
    Council (VR), the Swedish Foundation of Strategic Research (SSF),
    and the Excellence Center at Link\"{o}ping-Lund in Information
    Technology (ELLIIT). Parts of this work were presented at the IEEE
    International Conference on Acoustics, Speech, and Signal
    Processing (ICASSP) 2012. The authors believe in reproducible
    research and will upload the C++ code as supplementary material
    (media) on IEEEXplore.}}

\title{SUMIS: Near-Optimal Soft-In Soft-Out MIMO Detection With Low
  and Fixed Complexity}

\begin{document}
\maketitle

\begin{abstract}
  The fundamental problem of our interest here is soft-input
  soft-output multiple-input multiple-output (MIMO) detection. We
  propose a method, referred to as subspace marginalization with
  interference suppression (SUMIS), that yields unprecedented
  performance at low and fixed (deterministic) complexity. Our method
  provides a well-defined tradeoff between computational complexity
  and performance. Apart from an initial sorting step consisting of
  selecting channel-matrix columns, the algorithm involves no
  searching nor algorithmic branching; hence the algorithm has a
  completely predictable run-time and allows for a highly parallel
  implementation. We numerically assess the performance of SUMIS in
  different practical settings: full/partial channel state
  information, sequential/iterative decoding, and low/high rate outer
  codes. We also comment on how the SUMIS method performs in systems
  with a large number of transmit antennas.
\end{abstract}

%\linenumbers
\section{Introduction}
\label{sec:intro}
\thispagestyle{empty}

%\linenumbers

We consider multiple-input multiple-output (MIMO) systems, which are
known to have high spectral efficiency in rich scattering environments
\cite{Telatar} and high link robustness.  A major difficulty in the
implementation of MIMO systems is the detection (signal separation)
problem, which is generally computationally expensive to solve. This
problem can be especially pronounced in large MIMO systems, see
\cite{RusekPersson,Srinidhi,SomDatta,DattaSrinidhi,LiMurch,DattaKumar}
and the references therein.  The complexity of the optimal detector,
which computes the log-likelihood ratio (LLR) values exactly and
therefore solves the MIMO detection problem optimally, grows
exponentially with the number of transmit antennas and polynomially
with the size of the signal constellation. The main difficulty in MIMO
detection is the occurrence of ill-conditioned channels.  Suboptimal
and fast methods, such as zero-forcing, perform well only for
well-conditioned channels. With coded transmission, if the channel is
ill-conditioned and there is no coding across multiple channel
realizations, using the zero-forcing detector can result in large
packet error probabilities. Dealing with ill-conditioned channels is
difficult and requires sophisticated techniques.

Many different methods have been proposed during the past two decades
that aim to achieve, with reduced computational complexity, the
performance of the optimal detector \cite{Larsson, BaiChoiBook,
  WubbenSeethaler, Elkhazin, LarssonJalden, BarberoThompson, ChoiShim,
  ViterboBoutros, GuoNilsson, Papa, StuderBolcskei,ChoiSinger}. A concise
exposition of some state-of-the-art MIMO detection technique can be
found in \cite{Larsson} and a more extensive overview is given in
\cite{BaiChoiBook}. Detection methods based on lattice reduction are
explained in some detail in \cite{WubbenSeethaler}. Most of today's
state-of-the-art detectors provide the possibility of trading
complexity for performance via the choice of some user parameter. One
important advantage of such detectors is that this parameter can be
adaptively chosen depending on the channel conditions, in order to
improve the overall performance \cite{Cirkic,NikitopoulosAscheid}.

There are two main categories of MIMO detectors. The first consists of
detectors whose complexity (run-time) depends on the particular
channel realization.  This category includes, in particular, methods
that perform a tree-search. Notable examples include the
sphere-decoding (SD) aided max-log method and its many relatives
\cite{ViterboBoutros,GuoNilsson,ChoiShim,Papa,StuderBolcskei}. A
recent method in this category is the reduced dimension
maximum-likelihood search (RD-MLS)
\cite{ChoiShim,BaiChoiBook}. Unfortunately, these methods have an
exponential worst-case complexity unless a suboptimal termination
criterion is used. The other category of detectors consists of methods
that have a fixed (deterministic) complexity that does not depend on
the channel realization. These methods are more desirable from an
implementation point of view, as they eliminate the need for data
buffers and over-dimensioning (for the worst-case) of the
hardware. Examples of such detectors are the reduced dimension maximum
a posteriori (RDMAP) method \cite{Elkhazin}, the partial
marginalization (PM) method \cite{LarssonJalden}, and the
fixed-complexity SD (FCSD) aided max-log method
\cite{BarberoThompson}. These fixed-complexity detectors provide a
simple and well-defined tradeoff between computational complexity and
performance, they have a fixed and fully predictable run-time, and
they are highly parallelizable. We will discuss existing detectors in
more detail in Section~\ref{ssec:stateof}.

\paragraph*{Summary of  Contributions}

We propose a new method that is inspired by the ideas in
\cite{ChoiShim,LarssonJalden,BarberoThompson,Elkhazin} of partitioning
the original problem into smaller subproblems. As in the PM and RDMAP
methods, we perform marginalization over a few of the bits when
computing the LLR values. The approximate LLRs that enter the
marginalization are much simpler than those in PM, and this
substantially reduces the complexity of our algorithm, as will be
explained in more detail in \secref{sec:sumis}.  In addition, we
suppress the interference on the considered subproblems (subspaces) by
performing soft interference suppression (SIS). The core idea behind
SIS germinated in \cite{Taylor} in a different context than MIMO
detection, but the use of SIS as a constituent of our proposed
algorithm is mainly inspired by the work in
\cite{Taylor,WangPoor,Schniter,ChoiCheong}.  The main differences
between the SIS procedure used in our algorithm and that in
\cite{WangPoor,Schniter,ChoiCheong} are: (i) we allow for the signal
and interference subspaces to have varying dimensionality; (ii) we
perform the SIS in a MIMO setting internally without the need for a
priori information from the decoder, as opposed to
\cite{WangPoor,Elkhazin}; and (iii) we do not iterate the internal LLR
values more than once, nor do we ignore the correlation between the
interfering terms over the different receive antennas as in
\cite{Schniter,ChoiCheong}. We refer to our method as \emph{subspace
  marginalization with interference suppression} (SUMIS). During the
review of this paper, reference \cite{ChoiLee} was brought to our
attention. Reference \cite{ChoiLee}, published a year after SUMIS was
first presented \cite{CirkicLarsson}, discusses another variation on
the theme where the interference is suppressed successively in
contrast to SUMIS which does this in parallel.

SUMIS was developed with the primary objective not to require
iteration with the channel decoder as this increases latency and
complexity.  However, SUMIS takes soft input, so the overall decoding
performance can be improved by iterating with the decoder. The ideas
behind SUMIS are fundamentally simple and allow for massively parallel
algorithmic implementations.  As demonstrated in
Section~\ref{sec:numres}, the computational complexity of SUMIS is
extremely low, and the accuracy is near or better than that of
max-log. SUMIS works well for both under- and over-determined MIMO
systems.

This paper extends our conference paper \cite{CirkicLarsson} by
including: a detailed complexity analysis, techniques for exploiting
soft input (non-uniform a priori probabilities), and techniques for
dealing with higher-order constellations and imperfect channel state
information. In addition, we discuss and exemplify the applicability
of SUMIS to systems with a large number of antennas.

\section{Preliminaries}
\label{sec:prel}

We consider the real-valued MIMO-channel model
\begin{equation}
  \label{eq:model}
  \bsy=\bsH\bss+\bse,
\end{equation}
where $\bsH\in\bbR^{\Nr\times\Nt}$ is the MIMO channel matrix and
$\bss\in\calS^{\Nt}$ is the transmitted vector.  We assume that
$\calS=\{-1,+1\}$ is a binary phase-shift keying (BPSK) constellation,
hence referring to a ``symbol'' is equivalent to referring to a
``bit''. With some extra expense of notation, as will be clear later,
it is straightforward to extend all results that we present to higher
order constellations. Furthermore, $\bse\in\bbR^{\Nr}\sim\calN%
(\bszero,\frac{\No}{2}\bsI)$ denotes the noise vector and
$\bsy\in\bbR^{\Nr}$ is the received vector. The channel is perfectly
known to the receiver unless stated otherwise. Also, hereinafter we
assume that $\Nr\geq\Nt$ since this is typical in practice and
simplifies the mathematics in the paper. Note that unlike many
competing methods, SUMIS does not require $\Nr\geq\Nt$---but this
assumption is made to render the comparisons fair.

Throughout we think of (\ref{eq:model}) as an \emph{effective channel
  model} for the MIMO transmission. In particular, if pure spatial
multiplexing is used, the matrix $\bsH$ in (\ref{eq:model}) just
comprises the channel gains between all pairs of transmit and receive
antennas.  If linear precoding is used at the transmitter, then $\bsH$
represents the combined effect of the precoding and the propagation
channel. In the latter case, ``transmit antennas'' should be
interpreted as ``simultaneously transmitted streams''.

Note that with separable complex symbol constellations (quadrature
amplitude modulation), every complex-valued model
\def\cx{{\text{c}}}
\begin{align}
\label{eq:cxmodel}
\bsy_\cx&=\bsH_\cx\bss_\cx+\bse_\cx,&\bse_\cx&\sim\calC\calN(0,\No\bsI),
\end{align}
where $(\cdot)_\cx$ denotes the complex-valued counterparts of
\eqref{eq:model}, can be posed as a real-valued model on the form 
\eqref{eq:model} by setting
\begin{subequations}
  \label{eq:cxtorlmodel}
  \begin{align}
    \hfill\bsy&=\begin{bmatrix}\Re{\bsy_\cx}\\\Im{\bsy_\cx}\end{bmatrix},&
    \hfill\bss&=\begin{bmatrix}\Re{\bss_\cx}\\\Im{\bss_\cx}\end{bmatrix},&
    \hfill\bse&=\begin{bmatrix}\Re{\bse_\cx}\\\Im{\bse_\cx}\end{bmatrix},
  \end{align}
  and 
  \begin{equation}
    \bsH=\begin{bmatrix}\Re{\bsH_\cx}&-\Im{\bsH_\cx}\\\Im{\bsH_\cx}&\Re{\bsH_\cx}\end{bmatrix}.
  \end{equation}
\end{subequations}
We use the real-valued model throughout as we will later be
partitioning $\bsH$ into submatrices, and then the real-valued model
offers some more flexibility: selecting one column in the
representation \eqref{eq:cxmodel} means simultaneously selecting two
columns in the representation \eqref{eq:model} (via
\eqref{eq:cxtorlmodel}), which is more restrictive. Simulation
results, not presented here due to lack of space, confirmed this with
a performance advantage in working with the real-valued
model. However, the difference is not major in most relevant
cases. Disregarding this technicality, it is straightforward to
re-derive all results in the paper using a complex-valued model
instead. That could be useful, for example if $M$-ary phase-shift
keying or some other non-separable signal constellation is used per
antenna.

\subsection{Optimal Soft MIMO Detection}

The optimal soft information desired by the channel decoder is the a
posteriori log-likelihood ratio
$l(s_i|\bsy)\triangleq\log\big(\frac{P(s_i=+1|\bsy)}{P(s_i=-1|\bsy)}\big)$
where $s_i$ is the $i$th bit of the transmitted vector $\bss$. The
quantity $l(s_i|\bsy)$ tells us how likely it is that the $i$th bit of
$\bss$ is equal to minus or plus one, respectively. By marginalizing
out all bits except the $i$th bit in $P(\bss|\bsy)$ and using Bayes'
rule, the LLR becomes
\begin{equation}
\label{eq:apostllr}
\begin{split}
l(s_i|\bsy)&=\log\left(\dfrac{\sum_{\bss:s_i(\bss)=+1}P(\bss|\bsy)}{\sum_{\bss:s_i(\bss)=-1}P(\bss|\bsy)}\right)\\
&=\log\left(\dfrac{\sum_{\bss:s_i(\bss)=+1}p(\bsy|\bss)P(\bss)}{\sum_{\bss:s_i(\bss)=-1}p(\bsy|\bss)P(\bss)}\right),
\end{split}
\end{equation}
where the notation $\sum_{\bss:s_i(\bss)=x}$ means the sum over all
possible  vectors $\bss\in\calS^\Nt$ for which the $i$th bit is
equal to $x$. With uniform a priori probabilities, i.e.,
$P(\bss)=1/2^\Nt$, the LLR can be written as
\begin{equation}
\label{eq:llr}
\begin{split}
l(s_i|\bsy)&=\log\left(\dfrac{\sum_{\bss:s_i(\bss)=+1}p(\bsy|\bss)}{\sum_{\bss:s_i(\bss)=-1}p(\bsy|\bss)}\right)\\
&=\log\left(\frac{\sum_{\bss:s_i(\bss)=+1}\exp{-\frac{1}{\No}\norm{\bsy-\bsH\bss}^2}}%
  {\sum_{\bss:s_i(\bss)=-1}\exp{-\frac{1}{\No}\norm{\bsy-\bsH\bss}^2}}\right).%
\end{split}
\end{equation}

In \eqref{eq:apostllr} and \eqref{eq:llr}, there are $2^\Nt$ terms
that need to be evaluated and added. The complexity of this task is
exponential in $\Nt$ and this is what makes MIMO detection
difficult. Thus, many approximate methods have been proposed. One very
good approximation of \eqref{eq:llr} is the so called max-log
approximation \cite{Robertson},
\begin{equation}
\label{eq:maxlog}
\begin{split}
l(s_i|\bsy)\approx\log\left(\frac{\max_{\bss:s_i(\bss)=+1}\exp{-\frac{1}{\No}\norm{\bsy-\bsH\bss}^2}}%
  {\max_{\bss:s_i(\bss)=-1}\exp{-\frac{1}{\No}\norm{\bsy-\bsH\bss}^2}}\right),%
\end{split}
\end{equation}
where only the largest terms in each sum of \eqref{eq:llr} are
retained, i.e, the terms for which $\norm{\bsy-\bsH\bss}$ is as small
as possible.  Typically, for numerical stability, sums as in
(\ref{eq:llr}) are evaluated by repeated use of the Jacobian
logarithm: $\log(e^a+e^b)=\mbox{max}(a,b)+\log(1+e^{-|a-b|})$, where
the second term can be tabulated as function of $|a-b|$. Max-log can
then be viewed as a special case where the second term in the Jacobian
logarithm is neglected.  Note that even though max-log avoids the
summation, one needs to search over $2^\Nt$ terms to find the largest
ones; hence the exponential complexity remains. Nevertheless, with the
max-log approximation, one can make any hard decision detector, such
as SD, to produce soft values. This has resulted in much of the
literature focusing on finding efficient hard decision methods. In
this paper, we make a clean break with this philosophy and instead
devise a good approximation of the LLRs \eqref{eq:apostllr} and
\eqref{eq:llr} directly.

In order to explain our proposed method and the competing
state-of-the-art methods, for fixed $\ns\in\{1,\dots,\Nt\}$, we define
the following partitioning of the model in \eqref{eq:model}
\begin{equation}
\label{eq:partmodel}
\bsy=\bsH\bss+\bse=\!\!\underbrace{\left[\wbH\quad\wtH\right]}%
_{\text{col. permut. of }\bsH}%
\text{\makebox{$\underbrace{\left[\bar{\bss}^T\;\;\tilde{\bss}^T\right]^T}%
_{\text{permut. of }\bss}$}}\!\!+\;\bse%
=\wbH\bar{\bss}+\wtH\tilde{\bss}+\bse,
\end{equation}
where $\wbH\in\bbR^{\Nr\times{\ns}}$,
$\wtH\in\bbR^{\Nr\times(\Nt-\ns)}$, $\bar{\bss}\in\calS^{\ns}$
contains the $i$th bit $s_i$ in the original vector $\bss$, and
$\tilde{\bss}\in\calS^{\Nt-\ns}$. The choice of partitioning involves
the choice of a permutation, and how to make this choice (for $\ns>1$)
is not obvious. For each bit in $\bss$, there are
$\binom{\Nt-1}{\ns-1}$ possible permutations in
\eqref{eq:partmodel}. How we perform this partitioning is explained in
\secref{ssec:sumisperms}. Note that for different detectors, the
choice of partitioning serves different purposes.

\subsection{Today's State-of-the-Art MIMO Detectors}
\label{ssec:stateof}

Note that all of the methods explained in this subsection, except
for PM, are designed to deliver hard decisions. These methods can then
produce soft decisions by using the max-log approximation. 

\paragraph*{The PM Method} 
PM \cite{LarssonJalden} offers a tradeoff between exact and
approximate computation of \eqref{eq:llr}, via a parameter
$r=\ns-1\in\{0,\dots,\Nt-1\}$.  We present the slightly modified
version in \cite{PerssonLarsson} of the original method in
\cite{LarssonJalden}, which is simpler than that in
\cite{LarssonJalden} but without comprising performance. The PM method
implements a two-step approximation of \eqref{eq:llr}. More
specifically, in the first step it approximates the sums of
\eqref{eq:llr} that correspond to $\tilde{\bss}\in\calS^{\Nt-\ns}$
with a maximization,
\begin{equation}
  \label{eq:pmmaxlog}
  l(s_i|\bsy)\approx%
  \log\left(\frac%
           {\displaystyle\sum_{\text{{\makebox[10mm]{$\quad\bar{\bss}\!\!:\!s_i\!(\!\bss)\!=\!+1$}}}}%
             \text{\raisebox{1pt}{$\hspace{-3mm}\displaystyle\max_{\tilde{\bss}}%
             \expp\Big(\!\!-\!\text{\makebox{\footnotesize$\frac{1}{\No}$}}\,%
             \Vert\bsy-\wbH\bar{\bss}-\wtH\tilde{\bss}\Vert^2\Big)$}}}
           {\text{\raisebox{-7pt}{$\displaystyle\sum_{\text{{\makebox[11mm]{$\quad\bar{\bss}\!\!:\!s_i\!(\!\bss)\!=\!-1$}}}}$}%
             \raisebox{-6pt}{$\hspace{-3mm}\displaystyle\max_{\tilde{\bss}}%
             \expp\Big(\!\!-\!\text{\footnotesize\makebox{$\frac{1}{\No}$}}\,%
             \Vert\bsy-\wbH\bar{\bss}-\wtH\tilde{\bss}\Vert^2\Big)$}}}\right).
\end{equation}
In the second step, the maximization in \eqref{eq:pmmaxlog} is
approximated with a linear filter with quantization (clipping), such
as the zero-forcing with decision-feedback (ZF-DF) detector
\cite{LarssonJalden}. The ZF-DF method is computationally much more
efficient than exact maximization, but it performs well only for
well-conditioned matrices $\wtH$. However, the $\max$ problems in
\eqref{eq:pmmaxlog} are generally well-conditioned since the matrices
$\wtH$ are typically tall. For PM, when forming the partitioning in
\eqref{eq:partmodel}, the original bit-order in
$\bss=[s_1,\dots,s_\Nt]^T$ is permuted in \eqref{eq:pmmaxlog} in a way
such that the condition number of $\wtH$ is minimized, see
\cite{LarssonJalden}. Notably, PM performs ZF-DF aided max-log
detection in the special case of $r=0$ and computes the exact LLR
values (as defined by \eqref{eq:llr}) for $r=\Nt-1$.

\paragraph*{The FCSD Method}
As already noted, SD is a well known method for computing hard
decisions.  All variants of SD have a random complexity (runtime).
This is a very undesirable property from a hardware implementation
point of view, and this is one of the insights that stimulated the
development of the FCSD method \cite{BarberoThompson} (as well as PM
\cite{LarssonJalden} and our proposed SUMIS). FCSD performs
essentially the same procedure as the PM method except that it
introduces an additional approximation by employing max-log on the
sums in the PM method, i.e., the sums over
$\{\bar{\bss}\in\calS^{\ns}:s_i(\bss)=x\}$ in
\eqref{eq:pmmaxlog}. Hence, instead of summing over
$\{\bar{\bss}\in\calS^{\ns}:s_i(\bss)=x\}$ for each $x$ as in PM, it
picks the best candidate from
$\{\bar{\bss}\in\calS^{\ns}:s_i(\bss)=x\}$ for each $x$. As a result,
the FCSD method offers a tradeoff between exact and approximate
computation of the max-log problem in \eqref{eq:maxlog}.

\paragraph*{The SD Method and Its Soft-Output Derivatives}

The conventional SD method \cite{Murugan}, which also constitutes the
core of its many derivatives such as
\cite{WangGiannakis,ChoiShim,BaiChoiBook,StuderBolcskei,Mennenga,HochwaldBrink},
does not use the partitioned model in \eqref{eq:partmodel}. It uses a
(sorted) QR-decomposition $\bsH=\bsPi\bsR$, where
$\bsPi\in\bbR^{\Nr\times\Nt}$ has orthonormal columns and
$\bsR\in\bbR^{\Nt\times\Nt}$ is an upper-triangular matrix, in order
to write $\argmin_{\bss:s_i=x}\big\lVert\bsy-\bsH\bss\big\rVert^2=%
\argmin_{\bss:s_i=x}\big\lVert\bsPi^T\bsy-\bsR\bss\big\rVert^2$ for
each $x$ so that the max-log problem in \eqref{eq:maxlog} can be
solved by means of a tree-search. This search requires the choice of
an initial sphere radius and success is only guaranteed if the initial
radius is large enough and if the search is not prematurely terminated
\cite{Murugan}. SD has a high complexity for some channel realizations
(the average complexity is known to be exponential in $\Nt$
\cite{JaldenOttersten}), and hence, in practice a stopping criterion
is used to terminate the algorithm after a given number of
operations. Particularly, SD requires more time to finish for
ill-conditioned than for well-conditioned channel realizations. To
reduce the negative effect of ill-conditioned matrices, matrix
regularization \cite{Rugini,StuderBolcskei} or lattice reduction
techniques \cite{WubbenSeethaler} can be applied. The idea of the
latter is to apply, after relaxing the boundaries of the signal
constellation, a transformation that effectively reduces the condition
number of the channel matrix. Using SD to solve \eqref{eq:maxlog}
naively would require two runs of the procedure per bit, which becomes
computationally prohibitive when detecting a vector of bits. We next
review some methods that tackle this issue.

In the list sphere-decoder \cite{HochwaldBrink}, a list of a fixed
number of candidates is stored during the search through the tree (one
single tree-search is performed). Then \eqref{eq:maxlog} is solved
approximately by picking out the minimum-norm candidates in the
resulting candidate list instead of in the full multi-dimensional
constellation. Unfortunately, this procedure does not guarantee that
the signal vector candidates with the smallest norms are included in
the resulting list and thus performance is compromised. The list size
and the sphere radius must be very carefully chosen. A more
sophisticated version of this method can be found in \cite{Mennenga}.

In the repeated tree-search method \cite{WangGiannakis}, the
tree-search is performed multiple times but not as many times as in
the naive approach. In the first run, the algorithm performs one
single tree-search that finds the vector with the smallest norm
$\argmin_{\bss}\big\lVert\bsPi^T\bsy-\bsR\bss\big\rVert^2$. This in
effect finds
$\argmin_{\bss:s_i=x}\big\lVert\bsPi^T\bsy-\bsR\bss\big\rVert^2$ for
each bit but only for one of the bit hypotheses (either $x=+1$ or
$x=-1$). To find the smallest-norm candidates for the
counterhypothesis $x^\complement$, a tree-search is performed for each
bit to solve
$\argmin_{\bss:s_i=x^\complement}\big\lVert\bsPi^T\bsy-\bsR\bss\big\rVert^2$. This
method finds the minimums in \eqref{eq:maxlog} using only half the
number of tree-searches required by the naive approach. In this
method, the main disadvantage remains, which is that it may visit the
same nodes multiple times.

The single-tree-search method \cite{StuderBolcskei}, on the other hand,
traverses the multiple trees in parallel instead of in a repeated
fashion as is done in \cite{WangGiannakis}. While doing that, it keeps
track of which nodes have been already visited in one search so that
they can be skipped in the rest. That way, this method can guarantee
to find the minimums in \eqref{eq:maxlog} with one extended single
tree-search and hence provides clear advantages over methods in both
\cite{HochwaldBrink} and \cite{WangGiannakis}.

\paragraph*{The RD-MLS Method}
RD-MLS \cite{ChoiShim,BaiChoiBook} carries out the same procedure as
FCSD except that it does not perform clipping after the linear
filtering. Instead, it uses an SD type of algorithm to perform a
reduced tree-search over $\{\bar{\bss}\in\calS^{\ns}:s_i(\bss)=x\}$
for each $x$. Although this method reduces the number of layers in the
tree, it does not necessarily improve the conditioning of the reduced
problem, as the PM and FCSD methods do. This is so due to the
unquantized linear filtering operation that in effect results in
performing a projection of the original space (column space of $\bsH$)
onto the orthogonal complement of the column space of
$\wtH$. Therefore, for an ill-conditioned matrix $\bsH$, even though
RD-MLS searches over a reduced space (roughly half the original space
dimension \cite[Sec. V]{ChoiShim}, i.e., $\ns\approx\Nt/2$), it is
unclear whether the RD-MLS algorithm would visit significantly fewer
nodes in the reduced space $\bar{\bss}$ than in the original space
$\bss$. The reason is that the RD-MLS will suffer from the same
problem in the reduced space as the conventional SD would have in the
original space, namely the slow reduction in radius and pruning of
nodes that are not of interest.

\section{Proposed Soft MIMO Detector (SUMIS)}
\label{sec:sumis}

Our proposed method, SUMIS, consists of two main stages. In stage I, a
first approximation to the a posteriori probability of each bit $s_i$
is computed. In stage II, these approximate LLRs are used in an
interference suppression mechanism, whereafter the LLR values are
calculated based on the resulting ``purified'' model.  To keep the
exposition simple, we herein first present a the basic ideas behind
SUMIS for the case that $P(\bss)$ is uniform. The extension to
non-uniform $P(\bss)$ is treated in \secref{sec:nuniprob}. A highly
optimized version (which has much lower complexity but sacrifices
clarity of exposition) of SUMIS is then presented in
\appref{app:cmplx}. A practical implementation of SUMIS should use the
version in \appref{app:cmplx}.

\subsection{Stage I} We start with the partitioned model in
\eqref{eq:partmodel}
\begin{equation}
\label{eq:subsmodel}
\bsy=\wbH\bar{\bss}\;+\;\underbrace{\wtH\tilde{\bss}\;+\;\bse}_{\makebox[0mm]{\scriptsize{interference+noise}}}
\end{equation}
and define an approximate model
$\bar{\bsy}\triangleq\wbH\bar{\bss}+\bsn$ where $\bsn$ is a Gaussian
stochastic vector $\calN(\bszero,\bsQ)$ with
$\bsQ\triangleq\wtH\wtPsi\wtH^T\!\!\!+\!\frac{\No}{2}\bsI$ and
$\wtPsi$ is the covariance matrix of $\tilde{\bss}$. Under the
assumption that the symbols are independent, $\wtPsi$ is diagonal, and
since $P(\bss)$ is uniform, $\wtPsi=\bsI$. It is important to note
that $\bar{\bsy}=\wbH\bar{\bss}+\bsn$ is an approximated model of
\eqref{eq:subsmodel}, which we will use to approximate the probability
density function $p(\bsy|\bar{\bss})\approx{}%
p(\bar{\bsy}|\bar{\bss})\big|_{\bar{\bsy}=\bsy}$. The approximation
consists of considering the interfering terms $\wtH\tilde{\bss}$ as
Gaussian. This is a reasonable approximation since each element in
$\wtH\tilde{\bss}$ constitutes a sum of variates and thus generally
has Gaussian behaviour, especially when the variates are many and
independent. This is a consequence of the central limit theorem, see
\cite[sec. 8.5]{Papoulis} and \cite[fig. 8.4b]{Papoulis}.

To compute the a posteriori probability $P(s_k|\bsy)$ of a bit $s_k$,
which is contained in $\bar{\bss}$, we can marginalize out the
remaining bits in $P(\bar{\bss}|\bsy)$. Note that computing
$P(\bar{\bss}|\bsy)$ itself requires marginalizing out $\tilde{\bss}$
from $P(\bss|\bsy)$, which is computationally very
burdensome. However, with our proposed  approximation, we can write
\begin{equation}
P(\bar{\bss}|\bsy)\propto{p}(\bsy|\bar{\bss})P(\bar{\bss})\propto{}%
p(\bsy|\bar{\bss})\approx{}p(\bar{\bsy}|\bar{\bss})\big|_{\bar{\bsy}=\bsy},
\end{equation}
and therefore approximate the a posteriori probability function
$P(s_k|\bsy)$ with the function
\begin{equation}
\label{eq:skprob}
P(s_k=s|\bar{\bsy})\big|_{\bar{\bsy}=\bsy}\propto%
\sum_{\bar{\bss}:s_k=s}p(\bar{\bsy}|\bar{\bss})\big|_{\bar{\bsy}=\bsy}.
\end{equation}
Note that the number of terms in the summation over $\bar{\bss}:s_k=s$
is $2^{\ns-1}$, which is significantly smaller than the number of
terms that need to be added when evaluating $P(s_k|\bsy)$ exactly.

Using the following operator
$\norm{\bsx}^2_\bsQ\triangleq\bsx^T\bsQ^{-1}\bsx$ for some vector
$\bsx\in\bbR^\Nr$, we write
\begin{equation}
\label{eq:prbybarsbar}
p(\bar{\bsy}|\bar{\bss})=\frac{1}{\sqrt{(2\pi)^\Nr|\bsQ|}}%
\exp{-\frac{1}{2}\norm{\bar{\bsy}-\wbH\bar{\bss}}^2_\bsQ}
\end{equation}
and due to the assumption on $\calS$ being BPSK, we can perform the
marginalization in \eqref{eq:skprob} in the LLR domain as (after
inserting $\bsy$ in place of $\bar{\bsy}$ in \eqref{eq:prbybarsbar})
\begin{equation}
\label{eq:skprobllr}
\lambda_k\triangleq\log\left(\frac%
    {\sum_{\bar{\bss}:s_k=+1}\exp{-\frac{1}{2}\norm{\bsy-\wbH\bar{\bss}}^2_\bsQ}}
    {\sum_{\bar{\bss}:s_k=-1}\exp{-\frac{1}{2}\norm{\bsy-\wbH\bar{\bss}}^2_\bsQ}}\right).
\end{equation}
The summations in \eqref{eq:skprobllr} can be computed efficiently,
with good numerical stability, and using low-resolution fixed-point
arithmetics via repeated use of the Jacobian logarithm. The a
posteriori probabilities of the remaining elements in $\bss$ are
approximated analogously to \eqref{eq:subsmodel}-\eqref{eq:skprobllr}
by simply choosing different partitionings (permutations) of $\bsH$
and $\bss$ such that the bit of interest is in $\bar{\bss}$. The main
purpose of stage I is to reduce the impact of the interfering
term $\wtH\tilde{\bss}$. For this purpose, we compute the conditional
expected value of bit $s_k$ approximately using the function
$P(s_k=s|\bar{\bsy})$,
\begin{equation}
\label{eq:skexpect}
\begin{split}
\Exp{s_k|\bsy}&\triangleq\sum_{s\in\calS}sP(s_k=s|\bsy)\approx\sum_{s\in\calS}sP(s_k=s|\bar{\bsy})\bigg|_{\bar{\bsy}=\bsy}\\
&=\frac{-1}{1+e^{\lambda_k}}+\frac{1}{1+e^{-\lambda_k}}=\tanh\big(\textstyle\frac{\lambda_k}{2}\big).
\end{split}
\end{equation}

Stage I is performed for all bits $s_k$ in $\bss$, i.e.,
$k=1,\dots,\Nt$. For higher-order constellations, this stage would be
performed symbol-wise rather than bit-wise as presented above. This is
the reason for using the index $k$ in place of index $i$ in the
description here. Note that the soft MMSE procedure computes
\eqref{eq:skprobllr} bit-wisely with $\ns$ set to one, i.e., soft MMSE
essentially is a special case of SUMIS when $\ns=1$ and stage II is
disabled. One can see this equivalence by fixing $\ns=1$ and comparing
the underlying approximation steps, from the exact LLR calculation, we
take prior to \eqref{eq:skprobllr} with those that soft MMSE takes.

\subsection{Stage II: Purification} 

For each bit $s_i$, the interfering vector
$\tilde{\bss}$ in \eqref{eq:subsmodel} is suppressed using
\begin{equation}
\label{eq:suppmodel}
\bsy'\triangleq\bsy-\wtH\Exp{\tilde{\bss}|\bsy}\!=\!\wbH\bar{\bss}+%
\underbrace{\wtH(\tilde{\bss}-\Exp{\tilde{\bss}|\bsy})+\bse}_{\text{interference+noise}}\approx\!\wbH\bar{\bss}+\bsn',
\end{equation}
where $\bsn'\sim\calN\big(\bszero,\bsQ'\big)$ with
$\bsQ'\triangleq\wtH\wtPhi\wtH^T\!\!\!+\!\frac{\No}{2}\bsI$ and
$\wtPhi$ is the conditional covariance matrix of $\tilde{\bss}$. Note
that it is natural to suppress the interference by subtracting its
conditional mean since this removes the bias that the interference
causes. Under the approximation that the elements in $\tilde{\bss}$
are independent conditioned on $\bsy$, we have that
\begin{equation}
\wtPhi=\Exp{\diag(\tilde{\bss})^2\big|\bsy}-\Exp{\diag(\tilde{\bss})\big|\bsy}^2,
\end{equation}
where the operator $\diag(\cdot)$ takes a vector of elements as input
and returns a diagonal matrix with these elements on its diagonal, and
the notation $\bsA^2$ means $\bsA\bsA$ for some square matrix
$\bsA$. Since $\calS=\{-1,+1\}$, we get
\begin{equation}
\wtPhi=\bsI-\diag(\Exp{\tilde{\bss}|\bsy})^2.
\end{equation}
After the interfering vector $\tilde{\bss}$ is suppressed and the
model is ``purified'', we compute the LLRs. The LLRs are computed by
performing a full-blown marginalization in \eqref{eq:llr} over the
corresponding $\ns$-dimensional subspace $\bar{\bss}$ using the
purified approximate model in \eqref{eq:suppmodel}. Hence, the LLR
value we compute for the $i$th bit is
\begin{equation}
\label{eq:sumisllr}
l(s_i|\bsy)\approx\log\left(\!\frac%
  {\sum_{\bar{\bss}:s_i(\bss)=+1}\exp{\!-\frac{1}{2}\norm{\bsy'-\wbH\bar{\bss}}^2_{\bsQ'}}}%
  {\sum_{\bar{\bss}:s_i(\bss)=-1}\exp{\!-\frac{1}{2}\norm{\bsy'-\wbH\bar{\bss}}^2_{\bsQ'}}}%
\!\right).
\end{equation}
For higher order constellations, this stage is performed bit-wise;
hence we use the index $i$.

\subsection{Choosing the Permutations} 
\label{ssec:sumisperms}

The optimal permutation that determines $\wbH$ and $\wtH$ would be the
one that minimizes the probability of a decoding error. This
permutation is hard to find as it is difficult to derive tractable
expressions of the probability of decoding error. There are many
possible heuristic ways to choose the permutations. For instance, in
the PM and FCSD methods, the aim is to find a permutation such that
the condition number, i.e., the ratio between the largest and smallest
singular values, of $\wtH$ is minimized. The reason for this choice is
that the matrix $\wtH$ determines the conditioning of the subproblem
in PM, which in turn is solved by applying a zero-forcing filter. In
SUMIS, by contrast, we aim to choose the partitioning such that for a
bit $s_k$ in $\bss$, the interfering vector $\tilde{\bss}$ in
\eqref{eq:subsmodel} has as little effect on the useful signal vector
$\bar{\bss}$ as possible. This in essence means that we would like the
interference to lie in the null-space of the useful signal, i.e., the
inner product between the columns in $\wbH$ and those in $\wtH$ should
be as small as possible. In the extreme case when the column spaces of
$\wbH$ and $\wtH$ are orthogonal, the marginalization over
$\bar{\bss}$ and $\tilde{\bss}$ decouples and SUMIS would become
optimal. In this respect, SUMIS fundamentally differs from PM and
FCSD, which instead would become optimal if the permutation could be
chosen so that $\wtH$ had orthognal columns.

We base our partitioning on $\bsH^T\bsH$, which has the structure
\begin{equation}
\bsH^T\bsH=\begin{bmatrix}\text{\footnotesize$\sigma_1^2$}&\text{\makebox[8mm]{\footnotesize$\rho_{1,2}$}}&\dots\\%
  \text{\makebox[8mm]{\footnotesize$\quad\rho_{1,2}$}}&\text{\footnotesize$\sigma_2^2$}&&\\\vdots&&\ddots\end{bmatrix}.
\end{equation}
For the $k$th column of $\bsH^T\bsH$ (corresponding to the $k$th bit
in $\bss$) we pick the $\ns\!-1$ indices that correspond to the
largest values of $|\rho_{k,\ell}|$. Then, these indexes along with
the index $k$ specify the columns from $\bsH$ that are placed in
$\wbH$. The rest of the columns are placed in $\wtH$. Therefore, the
choice of permutation will depend on $\wbH^T\wbH$ (the ``power'' of
$\wbH$) and the ``correlation'' $\wbH^T\wtH$. Note also that the
matrix $\bsH^T\bsH$ is used in the other SUMIS stages and therefore
evaluating it here does not add anything to the total computational
complexity. Hence, complexity-wise, computing the permutation
comes nearly for free. Also, as shown in the numerical results, using
the proposed permutation, SUMIS performs close to optimal.

\subsection{Summary} 
The steps of our SUMIS method are summarized in \algref{alg:sumis} in
the form of generic pseudo-code. Via the adjustable subspace
dimensionality, i.e., the parameter $\ns$, SUMIS provides a simple and
well-defined tradeoff between computational complexity and detection
performance. In the special case when $\ns=\Nt$, there is no
interfering vector $\tilde{\bss}$ and SUMIS performs exact LLR
computation. At the other extreme, if $\ns=1$ SUMIS becomes the soft
MMSE method with the additional step of model purification. The
complexity of SUMIS is derived in \appref{app:cmplx}, and the
complexity of some of its competitors is summarized in
\tabref{tab:complexity}. The complexity is measured in terms of
elementary operations bundled together: additions, subtractions,
multiplications, and divisions. Further, it is divided into two parts:
one part representing calculations done once for each channel matrix
(``$\bsy$-independent'') and one part to be done for each received
$\bsy$-vector (``$\bsy$-dependent''). We can see that PM requires
approximately $(2\Nt^3+4\Nt^2)2^\ns$ operations in the
$\bsy$-dependent part only, and max-log (via SD) requires
$3(\Nr+\Nt)\Nt^2$ in the initialization stage only\footnotemark. Thus, SUMIS
provides clear complexity savings and it does so, as we will see in
\secref{sec:numres}, at the same time as it offers significant
performance gains over these competing methods.

\footnotetext{The significant initialization complexity of SD is a
  result of the computation of the sorted QR decomposition
  \cite{WubbenBohnke} on regularized channel models
  \cite{Rugini,StuderBolcskei}. Channel regularization is used to reduce
  the impact of ill-conditioned channel matrices on the size of the
  tree-search. This complexity is much larger compared to the
  LDL-decomposition, which is used favorably in SUMIS, see
  \appref{app:cmplx}.}

\def\mytabspace{\parbox[b][3ex][b]{0mm}{}}
\begin{table*}
\centering
\begin{tabular}{|l|c|c|}
\hline
Det. method&$\bsy$-independent&$\bsy$-dependent\\
\hline
\hline
SUMIS (proposed)\mytabspace&$\Nr\Nt^2+\Nt^3+2\ns^2\Nt^2$&$\Nt^3+2\Nr\Nt+(2\ns^2+6)\Nt^2$\\
\hline
SD aided max-log&$3(\Nr+\Nt)\Nt^2$&\parbox[b][5ex][b]{5cm}{\centering{random and exponential in $\Nt$ on average \cite{JaldenOttersten}}}\\
\hline
PM\mytabspace&$\Nr\Nt^2+\Nt^3$&$(2\Nt^3+4\Nt^2)M^\ns$\\
\hline
soft MMSE\mytabspace&$\Nr\Nt^2+\Nt^3$&$2\Nt(\Nr+\Nt)$\\
\hline
max-log\mytabspace&$3\Nt{M}^\Nt$&$\Nt{M}^\Nt$\\
\hline
exact LLR\mytabspace&$3\Nt{M}^\Nt$&$\Nt{M}^\Nt$\\
\hline
\end{tabular}
\caption{Complexity summary in terms of elementary operations bundled
  together.  Here $M\triangleq|\calS|$ denotes the cardinality of the
  constellation.  The derivation of the SUMIS complexity is presented
  in \appref{app:cmplx}. PM includes initialization (same as for
  SUMIS) and the computation of $\Nt{M}^\ns$ norms required to produce
  the soft values. Soft MMSE requires a similar initialization
  (assuming that the matrix inverse is explicitly computed) and two
  matrix-vector multiplications to compute the required
  norms. SD-aided max-log includes a sorted QR-decomposition
  \cite{WubbenBohnke} of a regularized channel matrix
  \cite{Rugini,StuderBolcskei} during the initialization
  ($\bsy$-independent part) and a tree-search of random complexity per
  $\bsy$, see \figref{fig:snrflops} for an empirical example. For the
  max-log and exact LLR, we have included the complexity of computing
  all $M^\Nt$ norms and then performing the comparisons and
  summations, respectively. To achieve the presented complexity
  figures of max-log and exact LLR, we sort the symbol vectors such
  that each consecutive vector differs only in one element from the
  previous vector. This facilitates computation of all $2^\Nt$ norms
  via a simple differential update procedure. Note that this table
  presents the complexity figures of the bottlenecks of each algorithm
  and are representative for $\Nr\geq\Nt\gtrsim10$ and $\Nt\gg\ns$ (or
  $\Nt^2\gg{M}^\ns$ specifically for SUMIS, see
  \appref{app:cmplx:ydep}).}
\label{tab:complexity}
\end{table*}

\begin{algorithm}
\caption{SUMIS pseudo-code implementation}
\label{alg:sumis}
\begin{algorithmic}
  \STATE{Start with some $\bsH$, $\bsy$ and $\ns\in\{1,\dots,\Nt\}$}  
  \FOR{$k=1,\dots,\Nt$\makebox[0mm][l]{\hspace{6mm}\textit{// -- Stage I -- //}}} 
  \STATE{Decide upon a partitioning in \eqref{eq:partmodel} based on $\bsH^T\bsH$}
  \STATE{Calculate $\lambda_k$ in \eqref{eq:skprobllr} (cond. probability of $s_k$ in
    terms of LLR)}
  \STATE{Calculate $\Exp{s_k|\bsy}$ and
    $\Var{s_k|\bsy}=1-\Exp{s_k|\bsy}^2$ in \eqref{eq:skexpect}}
  \ENDFOR 
  \FOR{each bit in $\bss$\makebox[0mm][l]{\hspace{6mm}\textit{// --
        Stage II -- //}}} 
  \STATE{Suppress the interfering vector $\tilde{\bss}$ and calculate
  $\bsy'$ in \eqref{eq:suppmodel}}
  \STATE{Calculate the new covariance matrix $\bsQ'$}
  \STATE{Calculate the LLR in \eqref{eq:sumisllr}}
  \ENDFOR
\end{algorithmic}
\end{algorithm}

% Finally, to be able to compute the expectation value $\Exp{\bss}$,
% we need the probabilities $\Prob{\tilde{s}_j=\tilde{s}}$. They are
% computed by a product of probabilities of the corresponding bits in
% $\tilde{s}_j$, which are approximated using the corresponding
% so-obtained LLR values. Hence, let us say that one such bit in
% $\tilde{s}_j$ has the index $i_j$ in the original vector $\bss$,
% then the approximated probability of that bit being equal to zero is
% $1/(1\!+\!\exp{l(b_{i_j}|\bsy)})$ and being equal to one is
% $1\!-\!1/(1\!+\!\exp{l(b_{i_j}|\bsy)})$.

\section{Non-Uniform A Priori Probabilities}
\label{sec:nuniprob}

Algorithm \ref{alg:sumis} in \secref{sec:sumis} can be directly
extended to the case of non-uniform $P(\bss)$. The details for each
step are given as follows.

\subsection{Stage I}
Since we have a priori information on the symbols, we can purify the
model already in this stage and suppress the interfering subspace
$\tilde{\bss}$. First, we evaluate the expected value
$\Exp{s_k}\triangleq\sum_{s\in\calS}sP(s_k=s)$ and the purified
received data
\begin{equation}
  \bsy-\wtH\Exp{\tilde{\bss}}=\wbH\bar{\bss}+%
  \underbrace{\wtH(\tilde{\bss}-\Exp{\tilde{\bss}})+\bse}_{\text{interference+noise}},
\end{equation}
where the ``interference+noise'' is, as in \eqref{eq:subsmodel} in the
approximate model $\bar{\bsy}=\wbH\bar{\bss}+\bsn$, approximated to be
$\calN\big(\bszero,\bsQ\big)$ where now $\wtPsi$ in $\bsQ$ is not
necessarily equal to the identity matrix. More precisely, under the
restriction that $\calS=\{-1,+1\}$ and the assumption that the bits $s_k$
are independent, we get
\begin{equation}
\wtPsi=\bsI-\diag(\Exp{\tilde{\bss}})^2.
\end{equation}
We can approximate the a posteriori probability $P(s_k=s|\bsy)$,
analogously to \eqref{eq:skprob}, with
\begin{equation}
  \label{eq:skprobiter}
  P(s_k=s|\bar{\bsy})\propto\sum_{\bar{\bss}:s_k=s}p(\bar{\bsy}|\bar{\bss})P(\bar{\bss}).
\end{equation}
Using \eqref{eq:skprobiter}, we can approximate the expectation of
$s_k$ conditioned on $\bsy$ in the same manner as in
\eqref{eq:skexpect}, i.e.,
\begin{align}
  \label{eq:skexpectiter}
  &\Exp{s_k|\bsy}\approx\tanh\big(\textstyle\frac{\lambda_k}{2}\big),%
  &\lambda_k=\log\left(\frac{P(s_k=+1|\bar{\bsy})}{P(s_k=-1|\bar{\bsy})}\right)%
  \bigg|_{\bar{\bsy}=\bsy}.
\end{align}
Similarly to stage I in \secref{sec:sumis} for higher-order
constellations, this stage is performed symbol-wise; hence we use here
again the index $k$. Note that for BPSK constellations, the procedure
in \cite{Elkhazin} is equivalent to stage I here, i.e., the method of
\cite{Elkhazin} is a special case of SUMIS when stage II is disabled.

\subsection{Stage II} 

In this stage, exactly the same procedure is performed as in stage II
in \secref{sec:sumis} but with two minor modifications: first the model is
purified using \eqref{eq:skexpectiter} instead of \eqref{eq:skexpect},
and second, the LLR value of the $i$th bit is computed using
\begin{equation}
\label{eq:sumisllriter}
l(s_i|\bsy)\approx\log\left(\!\frac%
  {\sum_{\bar{\bss}:s_i(\bss)=+1}\exp{\!-\frac{1}{2}\norm{\bsy'\!-\!\wbH\bar{\bss}}^2_{\bsQ'}}P(\bar{\bss})}%
  {\sum_{\bar{\bss}:s_i(\bss)=-1}\exp{\!-\frac{1}{2}\norm{\bsy'\!-\!\wbH\bar{\bss}}^2_{\bsQ'}}P(\bar{\bss})}%
\!\right).
\end{equation}
For higher order constellations, this stage is performed bit-wise;
hence the index $i$.

\section{Imperfect Channel State Information}
\label{sec:icsi}

\def\Ntr{{N_{\text{TR}}}}
\def\Ntrvec{{1:\Ntr}}

In practice the receiver does not have perfect knowledge about $\bsH$.
Typically, the receiver then forms an estimate of the channel, based
on a known transmitted pilot matrix
$\bss^\Ntrvec\triangleq[\bss_1\dots\bss_\Ntr]$ and the corresponding
received matrix $\bsy^\Ntrvec\triangleq[\bsy_1\dots\bsy_\Ntr]$.  The
so-obtained estimated channel matrix will not be perfectly accurate,
and the estimation error should be taken into account when computing
the LLRs.

To optimally account for the imperfect knowledge of the channel,
$P(\bss|\bsy)$ (or $P(\bss|\bsy,\bsH)$ if the dependence on $\bsH$ is
spelled out explicitly) should be replaced with
$P(\bss|\bsy,\bsy^\Ntrvec,\bss^\Ntrvec)$ in \eqref{eq:apostllr}.  As
an approximation to the resulting optimal detector, instead of working
with $P(\bss|\bsy,\bsy^\Ntrvec,\bss^\Ntrvec)$, $\bsH$ may be replaced
with an estimate $\whH$ of $\bsH$ in $P(\bss|\bsy,\bsH)$. The
so-obtained detector is called \emph{mismatched}, and it is generally not
optimal, except for in some special cases \cite{TariccoBiglieri}.

We next extend SUMIS to take into account channel estimation
errors, i.e., using $P(\bss|\bsy,\bsy^\Ntrvec,\bss^\Ntrvec)$ in
\eqref{eq:apostllr}.  We model the channel estimate $\whH$, when
obtained using the training data $\bss^\Ntrvec$ and $\bsy^\Ntrvec$,
with
\begin{equation} \label{eq:chanest}
	\whH=\bsH+\bsDelta,
\end{equation}
where $\bsH$ is the true channel and $\bsDelta$ is the estimation
error matrix whose complex-valued counterpart $\bsDelta_\cx$ (which is
decomposed into $\bsDelta$ using \eqref{eq:cxtorlmodel}) has
independent $\calC\calN(0,\delta^2)$ elements where $\delta^2$ is the
variance of the estimation error per complex dimension. Typically,
$\delta^2$ is directly proportional to the noise variance $\No$. The
assumption that the elements in $\bsDelta$ are independent holds if
the pilots are orthogonal and the noise is uncorrelated. Further, we
assume that the elements in $\whH$ are independent of those in
$\bsDelta$; this is the case if MMSE channel estimation is used
\cite{KailathSayed}.  From \eqref{eq:model} and \eqref{eq:chanest}, we
have
\begin{equation}
  \label{eq:icsimodel}
  \bsy=\bsH\bss+\bse%
  =\whH\bss+\overbrace{\bse-\bsDelta\bss}^{\triangleq\bsepsilon}%
  =\whH\bss+\bsepsilon.
\end{equation}
For constellations that have constant modulus, i.e., satisfy say
$\norm{\bss}^2=\Nt\forall\bss$, we know that
$\bsepsilon\sim\calN(0,\frac{\Nt\delta^2+\No}{2}\bsI)$. In this case, SUMIS
can be directly applied to a modified data model where $\whH$ and
$\bsepsilon$ are the channel matrix and noise vector, respectively.

By contrast, for general signal constellations, $\norm{\bss}^2$ is not
equal to a constant, and the situation becomes more
complicated. Recall that $P(\bss|\bsy,\whH)=%
p(\bsy|\bss,\whH)P(\bss)/p(\bsy)$, where the goal is to approximate
$p(\bsy|\bss,\whH)$ via $p(\bsy|\bar{\bss},\whH)$ using the philosophy
of SUMIS. That is, we target $p(\bsy|\bar{\bss},\whH)$ by
approximating $\bsepsilon$ conditioned on $\bar{\bss}$ and $\whH$ as
Gaussian. Approximating $\bsepsilon\big|_{\bar{\bss},\scrwhH}$ as
Gaussian is reasonable since each element in $\bsepsilon$ consists of
a sum of independent variates. The covariance matrix of $\bsepsilon$
conditioned on $\bar{\bss}$ and $\whH$ is
$\frac{\big(\mathbb{E}\{\norm{\tilde{\bss}}^2\}+
  \norm{\bar{\bss}}^2\big)\delta^2+\No}{2}\bsI$. Thus, the power of
the effective noise is
$\frac{\big(\mathbb{E}\{\norm{\tilde{\bss}}^2\}+
  \norm{\bar{\bss}}^2\big)\delta^2+\No}{2}$ instead of $\frac{\No}{2}$
as in \secref{sec:sumis}. This power now depends on
$\mathbb{E}\{\norm{\tilde{\bss}}^2\}$ and $\bar{\bss}$, causing the
complexity of SUMIS to increase substantially. The reason is that the
inverse of $\bsQ$ must be recomputed for each permutation in
\eqref{eq:partmodel} and even more often for each $\bar{\bss}$; the
same applies in stage II of SUMIS for the $\bsQ'$ matrix.

To avoid this complexity increase, we introduce further
approximations. First, instead of $\mathbb{E}\{\norm{\tilde{\bss}}^2\}$, we
use
\begin{equation}
\eta\triangleq\sum_{\substack{\text{all}~\Nt\\\text{permuts.}}}%
\frac{1}{\Nt}\mathbb{E}\{\norm{\tilde{\bss}}^2\},
\end{equation}
where the sum is taken over all $\Nt$ permutations considered in SUMIS
for a particular $\bsH$. This is reasonable since $\Nt\gg\ns$ and at
most $\ns$ elements out of $\Nt-\ns$ in $\tilde{\bss}$ are replaced
from one permutation to another; hence,
$\mathbb{E}\{\norm{\tilde{\bss}}^2\}$ will not differ much over the
permutations. Second, using again that $\Nt\gg\ns$, the variations in
$\norm{\bar{\bss}}^2$ will have a minor effect on the absolute power
of the effective noise. Therefore, we replace $\norm{\bar{\bss}}^2$ by
$\frac{1}{|\calS|^\ns}\sum_{\bar{\bss}\in\calS^\ns}\norm{\bar{\bss}}^2=\ns$. Hence,
the simplified noise power becomes $\frac{(\eta+\ns)\delta^2+\No}{2}$,
which is constant for each stage and results in a SUMIS complexity
equivalent to that of the full CSI approach, i.e., the complexity
presented in \appref{app:cmplx}.

\section{Very Large MIMO Settings}
\label{sec:vlmimo}

SUMIS was developed mainly for moderately-sized MIMO systems but it is
also applicable to large MIMO channels.  Previous research on
detection in MIMO systems with a large number of transmit antennas has
focused on hard decision algorithms
\cite{Srinidhi,DattaSrinidhi,LiMurch,DattaKumar}.  The performance of
such algorithms when used in coded systems is always upper-bounded by
the performance of max-log. In some cases (see the numerical results
in \secref{sec:numres}), the performance loss of the max-log
approximation becomes larger when the number of antennas
increases. Hence, the philosophy behind SUMIS---to approximate the LLR
directly---seems to be beneficial.

Another interesting observation in the large-MIMO setting is that even
the soft MMSE method can potentially achieve the performance of the
exact LLR method and thus outperform the max-log method. This
observation is motivated by the central limit theorem and the
following argument. Consider the model in \eqref{eq:partmodel} for
$\ns=1$, which makes $\wbH$ a column vector, say $\bsh_i$, and
$\bar{\bss}$ a scalar, say $s_i$; hence,
$\bsy=\bsh_is_i+\wtH\tilde{\bss}+\bse$. We assume that the elements in
$\bss$ are independent, which is a very common assumption in the
literature and a very reasonable assumption in practice as the bits in
a codeword are typically interleaved. By the central limit theorem,
the distribution of $\bsy$ for a given $s_i$ will approach a Gaussian:
$\calN\big(\bsh_is_i,\wtH\wtPsi\wtH^T\!\!\!+\!\frac{\No}{2}\bsI\big)$,
as $\Nt$ grows \cite[sec. 8.5]{Papoulis}. The exact LLR, i.e.,
$L(s_i|\bsy)=\log(p(\bsy|s_i=+1))-\log(p(\bsy|s_i=-1))$, will then for
sufficiently large $\Nt$ look like the LLR function based on the
Gaussian distribution, i.e.,
\begin{equation}
\label{eq:llrvlmimo}
L(s_i|\bsy)\approx4\bsy^T(\wtH\wtPsi\wtH^T\!\!\!+\!\frac{\No}{2}\bsI)^{-1}\bsh_i,%
\quad\text{for}~\Nt\gg1.
\end{equation}
This is a simple but important observation which does not seem to have
been made in the existing literature on large-MIMO detection
\cite{RusekPersson,Srinidhi,SomDatta,DattaSrinidhi,LiMurch,DattaKumar}.
However, the question remains as to how large $\Nt$ needs to be for
the approximation in \eqref{eq:llrvlmimo} to be tight. According to
\cite[sec. 8.5]{Papoulis}, under some conditions, the approximation
becomes very tight even for small values of $\Nt$.  We are
particularly interested in determining how tight it is in terms of
frame-error rate for large but finite $\Nt$. Our investigation with
$\Nt=12$ and $\Nt=26$ indicates that the performance of the soft MMSE
method is much closer to that of the exact LLR method for larger
$\Nt$.  What is especially interesting for the SUMIS method is that
the performance gap between the soft MMSE method and the exact LLR
method is reduced remarkably via the procedure in stage II of
SUMIS. Recall that the SUMIS method performs the soft MMSE procedure
in stage I when $\ns=1$.

\section{Numerical Results}
\label{sec:numres}

\subsection{Simulation Setup}
\label{ssec:simsetup}

Using Monte Carlo simulations we evaluate the performance of SUMIS and
compare it to the performance of some competitors. Performance is
quantified in terms of frame-error rate (FER) as a function of the
normalized signal-to-noise ratio $\Eb/\No$ where $\Eb$ is the total
transmitted energy per information bit. To make the results
statistically reliable, we count $300$ frame errors for each simulated
point. We simulate $6\times6$ and $13\times13$ complex MIMO systems
with $M^2$-QAM where $M\in\{2,4\}$, which means that the detection is
performed on equivalent real-valued $12\times12$ and $26\times26$ MIMO
systems with 2-PAM and 4-PAM modulation. The channel is Rayleigh
fading where each complex-valued channel matrix element is
independently drawn from $\calC\calN(0,1)$. We use three different
highly optimized irregular low-density parity-check (LDPC) codes with
rates $\{2/9,1/2,5/6\}$ (note that $2/9\lesssim1/4$), each having a
codeword length of approximately $10000$ bits. We use the parameters
of \cite{LeeKim} for rate $1/4$ and
\cite{RichardsonUrbanke,RichardsonShokrollahi} for rate $1/2$. For
rate $5/6$, we use the technique in \cite{MacKay} to generate the code
parameters. For more details, see the uploaded supplementary material
to this article. Two different coherence times are used: slow fading
(each codeword sees one channel realization) and fast fading (each
codeword spans $40$ channel matrices), respectively. We plot the FER
curves of the exact LLR (as defined by \eqref{eq:llr}), the max-log
approximation \eqref{eq:maxlog}, SUMIS for $\ns=1,3$, SUMIS
stage-I-only (without the purification procedure) for $\ns=1,3$, and
PM for $\ns=r+1=3$ \cite{LarssonJalden}. We include the approximate
LLRs computed in stage I of SUMIS in order to show the performance
gain of the purification step (stage II) of SUMIS. Recall that SUMIS
stage-I-only, for $\ns=1$, is equivalent to the soft MMSE method. For
reasons of numerical stability and efficiency, all
log-of-sums-of-exponentials were evaluated via repeated use of the
Jacobian logarithm. The convention used in the figures that follow is
that dashed lines represent the proposed methods and the solid lines
represent the competing ones.

Since the FCSD method is an approximation of the PM method, we refrain
from plotting its performance curves. We also omit performance plots
of RD-MLS and all other SD-based methods due to the fact that they
approximate the max-log method, and thus they cannot perform
better. Also, the complexity of the initialization step of these
algorithms alone is higher than the complexity of the complete SUMIS
procedure, see \tabref{tab:complexity}. Moreover, their complexity
depends on the channel realization and they have a very high
complexity for some channel realizations (the average complexity is
exponential in $\Nt$ \cite{JaldenOttersten}).  Taken together, this
renders performance-complexity comparisons with SD-based methods less
interesting.  We have included the max-log curve as a universal
indicator of the performance that can be achieved by any SD-type
detector.

\subsection{Results}
\label{ssec:simres}

\begin{figure}[bt]  
  \psfrag{FER}[][][\psscale][0]{FER}
  \psfrag{SNR}[][][\psscale][0]{$\Eb/\No$ [dB]}
  \psfrag{llr}[][][\psscale][0]{exact LLR}
  \psfrag{lmax}[][][\psscale][0]{max-log}
  \psfrag{sumis}[][][\psscale][0]{\parbox{3cm}{\centering{SUMIS\\$\ns=1,3$}}}
  \psfrag{sumis1}[][][\psscale][0]{\parbox{3cm}{\centering{SUMIS\\stage-I-only\\$\ns=1,3$}}}
  \psfrag{spm}[][][\psscale][0]{\parbox{3cm}{\centering{PM $\ns=3$}}}
  \includegraphics[width=\columnwidth]{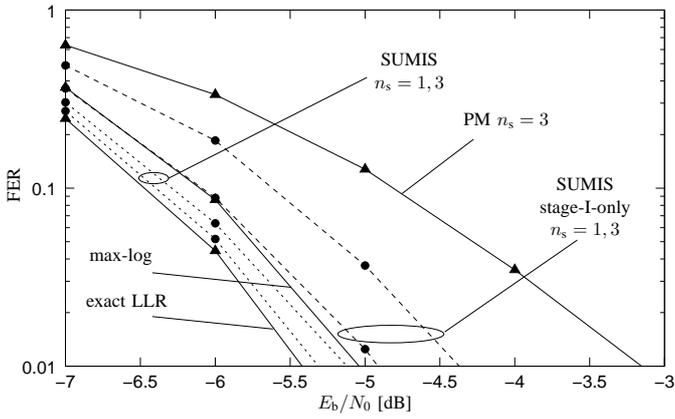}
    \caption{FER as a function of $\Eb/\No$ for the slow-fading
      $6\times6$ MIMO system ($\Nr=\Nt=12$ in \eqref{eq:model}) with
      $4$-QAM ($2$-PAM in \eqref{eq:model}) and with the LDPC code of
      rate $1/2$. The shown performance curves are: (i) dashed curves
      for the SUMIS stage-I-only and the complete SUMIS procedure with
      $\ns=1$ and $\ns=3$ spanning from right to left, and (ii) solid
      curves for the exact LLR method, the max-log method, and the PM
      method with $\ns=r+1=3$.}
\label{fig:6x6}
\end{figure}

\begin{figure}[bt]
   \psfrag{FER}[][][\psscale][0]{FER}
  \psfrag{SNR}[][][\psscale][0]{$\Eb/\No$ [dB]}
  \psfrag{llrsumis}[t][t][\psscale][0]{\parbox{3cm}{\centering{exact~LLR\\\vspace{-1mm}and~SUMIS}}}
  \psfrag{llrlmax}[t][t][\psscale][0]{\parbox{3cm}{\centering{exact~LLR\\\vspace{-1mm}and~max-log}}}
  \psfrag{lmax}[l][l][\psscale][0]{max-log}
  \psfrag{pm}[][][\psscale][0]{PM $\ns=3$}
  \psfrag{sumis}[b][b][\psscale][0]{\parbox{3cm}{\centering{SUMIS\\\vspace{-1mm}$\ns=1,3$}}}
  \psfrag{sumis1}[tc][tc][\psscale][0]{\parbox[][][s]{17mm}{SUMIS\vspace{-1mm}\\\mbox{}\hfill{stage-I-only}\hfill\mbox{}\vspace{-1mm}\\\mbox{}\hfill{$\ns=1,3$}}}
  \includegraphics[width=\columnwidth]{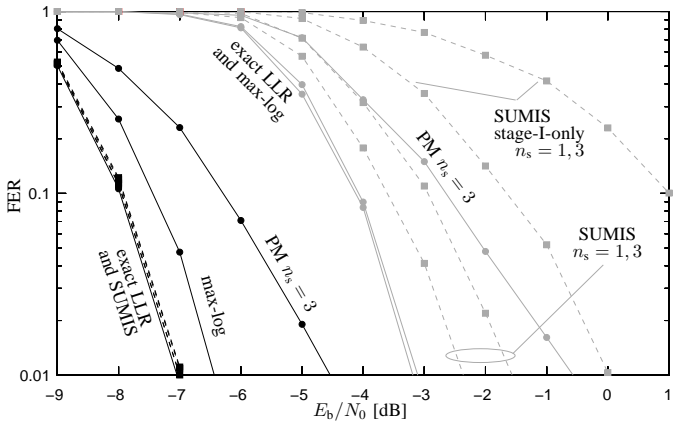}
  \caption{Same as in \figref{fig:6x6} but with LDPC codes of
    \textbf{rate} $\mathbf{2/9\lesssim1/4}$ \textbf{(black curves)} and
    \textbf{rate} $\mathbf{5/6}$ \textbf{(gray curves)}. Note that the
    left-most blended curves are five different curves: exact LLR,
    SUMIS for $\ns=1,3$, and SUMIS stage-I-only for $\ns=1,3$.}
\label{fig:6x6rates}
\end{figure}

\begin{figure}[bt]  
  \psfrag{FER}[][][\psscale][0]{FER}
  \psfrag{SNR}[][][\psscale][0]{$\Eb/\No$ [dB]}
  \psfrag{llr}[l][l][\psscale][0]{exact LLR}
  \psfrag{lmax}[l][l][\psscale][0]{max-log}
  \psfrag{sumis1}[l][l][\psscale][0]{\parbox{4cm}{(soft MMSE)\\SUMIS stage-I-only, $\ns=1$}}
  \psfrag{sumis}[l][l][\psscale][0]{SUMIS, $\ns=1$}
  \psfrag{pm}[][][\psscale][0]{PM, $\ns=3$}
  \includegraphics[width=\columnwidth]{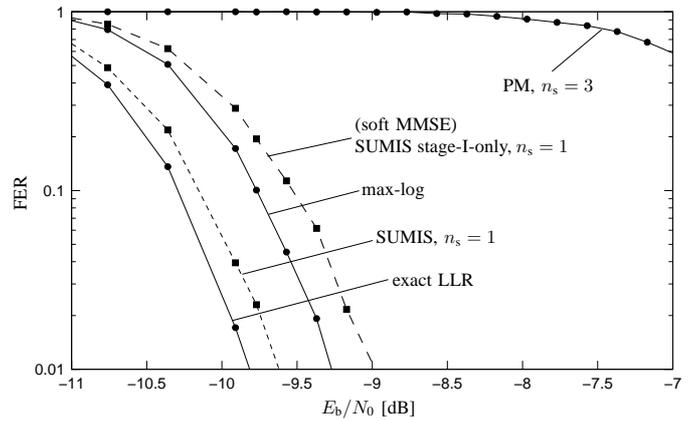}
  \caption{Same as in \figref{fig:6x6} but for a $13\times13$ MIMO
    system instead of $6\times6$. Note (in contrast to
    \figref{fig:6x6}) the increased gap between max-log and exact LLR,
    and the decreased gap between soft MMSE (SUMIS stage-I-only for
    $\ns=1$) and exact LLR. The SUMIS curve is very close to the exact
    LLR curve.}
\label{fig:13x13}
\end{figure}

\begin{figure}[bt]  
  \psfrag{FER}[][][\psscale][0]{FER}
  \psfrag{SNR}[][][\psscale][0]{$\Eb/\No$ [dB]}
  \psfrag{llr}[r][r][\psscale][0]{exact LLR}
  \psfrag{lmax}[r][r][\psscale][0]{max-log}
  \psfrag{sumis}[r][r][\psscale][0]{SUMIS, $\ns=1,3$}
  \psfrag{sumis1}[lt][lt][\psscale][0]{\parbox{3cm}{\centering{SUMIS stage-I-only,\\$\ns=1,3$}}}
  \psfrag{pm}[l][l][\psscale][0]{PM, $\ns=3$}
  \includegraphics[width=\columnwidth]{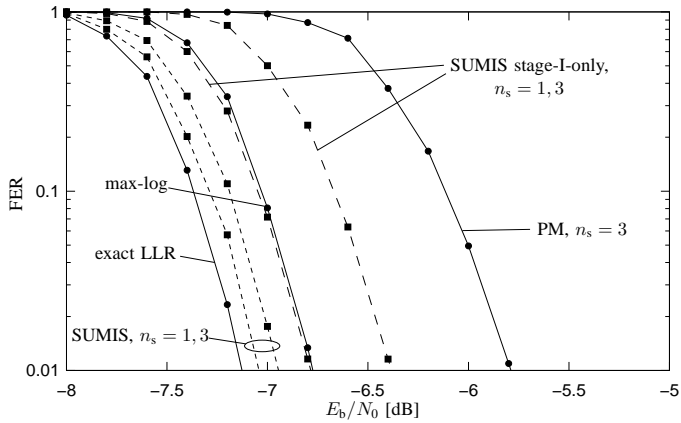}
  \caption{Same as in \figref{fig:6x6} but with fast-fading, i.e., a
    codeword spans $40$ independent channel realizations.}
\label{fig:6x6fast}
\end{figure}

\begin{figure}[bt]
  \psfrag{flops}[b][t][\psscale][0]{\#Elem. Ops. [$10^3$]}
  \psfrag{snr}[][][\psscale][90]{$\Eb/\No$ [dB]}
  \psfrag{llr}[t][t][\psscale][0]{exact LLR}
  \psfrag{maxlog}[b][b][\psscale][0]{\parbox[b][][b]{4cm}{\centering{max-log\\via full-search}}}
  \psfrag{min}[][][\psscale][0]{\texttt{min}}
  \psfrag{max}[][][\psscale][0]{\texttt{max}}
  \psfrag{mean}[][][\psscale][0]{\texttt{mean}}
  \psfrag{sdmaxlog}[b][b][\psscale][0]{\parbox[b][][b]{4cm}{\centering{SD aided max-log\\via single-tree-search \cite{StuderBolcskei}}}}
  \psfrag{sumis}[l][l][\psscale][0]{SUMIS}
  \psfrag{ns1}[r][r][\psscale][0]{$\ns=1$}
  \psfrag{ns3}[l][l][\psscale][0]{$\ns=3$}
  \psfrag{mmse}[l][l][\psscale][0]{soft MMSE}
  \psfrag{pm}[lb][lt][\psscale][0]{PM $\ns=3$}
  \includegraphics[width=\columnwidth]{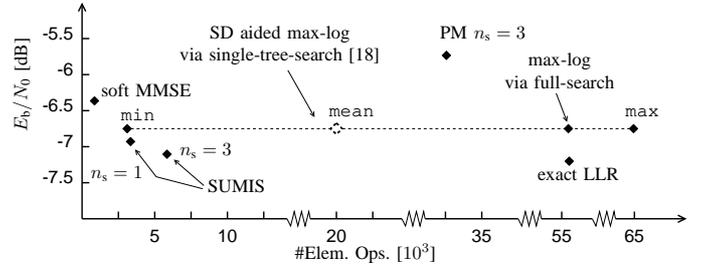}
\caption{Total complexity (both $\bsy$-independent and
  $\bsy$-dependent parts) per vector of bits in $\bss$ for the setting
  in \figref{fig:6x6fast}. In this setting, the channel stays the same
  for $21$ consecutive transmissions of $\bss$. The axes show the
  number of elementary operations versus the minimum signal-to-noise
  ratio required to achieve 1\% FER. For max-log via SD, we show the
  empirically evaluated minimum, maximum, and mean complexities.}
\label{fig:snrflops}
\end{figure}

\begin{figure}[bt]  
  \psfrag{FER}[][][\psscale][0]{FER}
  \psfrag{SNR}[][][\psscale][0]{$\Eb/\No$ [dB]}  
  \psfrag{lmax}[r][r][\psscale][0]{max-log}
  \psfrag{sumis}[r][r][\psscale][0]{SUMIS, $\ns=1,3$}
  \psfrag{sumis1}[][][\psscale][0]{SUMIS stage-I-only, $\ns=1,3$}
  \psfrag{pm}[r][r][\psscale][0]{PM, $\ns=3$}
  \includegraphics[width=\columnwidth]{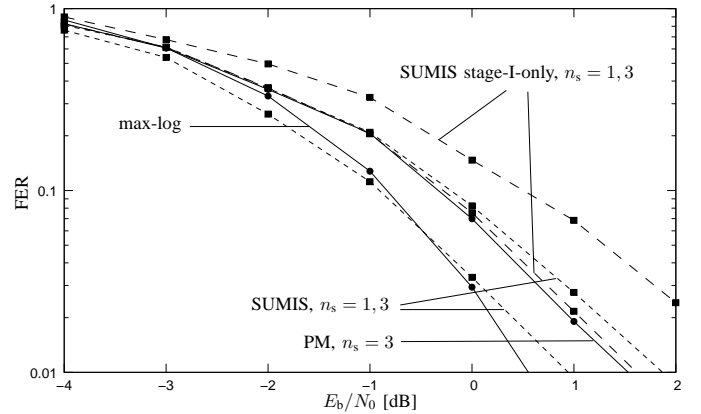}
  \caption{Same as in \figref{fig:6x6} but for $16$-QAM ($4$-PAM in
    \eqref{eq:model}) instead of $4$-QAM. The exact LLR curve has been
    excluded due to the massive complexity required to evaluate the
    FER. Its complexity is of the same order of magnitude as that in
    \figref{fig:13x13}.}
  \label{fig:6x6qam16}
\end{figure}

\begin{figure}[pbt]  
  \centering  
  \begin{subfigure}[bt]{\columnwidth}
    \psfrag{FER}[][][\psscale][0]{FER}
    \psfrag{SNR}[][][\psscale][0]{$\Eb/\No$ [dB]}
    \psfrag{llr}[][][\psscale][0]{LLR match.}
    \psfrag{mislmax}[l][l][\psscale][0]{max-log~mismatch.}
    \psfrag{lmaxmisllr}[][][\psscale][0]{\parbox{5cm}{\centering{max-log
          match.\\and LLR mismatch.}}}
    \psfrag{sumis}[][][\psscale][0]{\parbox{5cm}{\centering{SUMIS~$\ns=1,3$\\match.}}}
    \psfrag{missumis}[l][l][\psscale][0]{SUMIS~$\ns=1,3$~mismatch.}
    \psfrag{pm}[][][\psscale][0]{\parbox{5cm}{\centering{PM~$\ns=3$\\match.}}}
    \psfrag{mispm}[][][\psscale][0]{\parbox{5cm}{\centering{PM~$\ns=3$\\mismatch.}}}
    \includegraphics[width=\columnwidth]{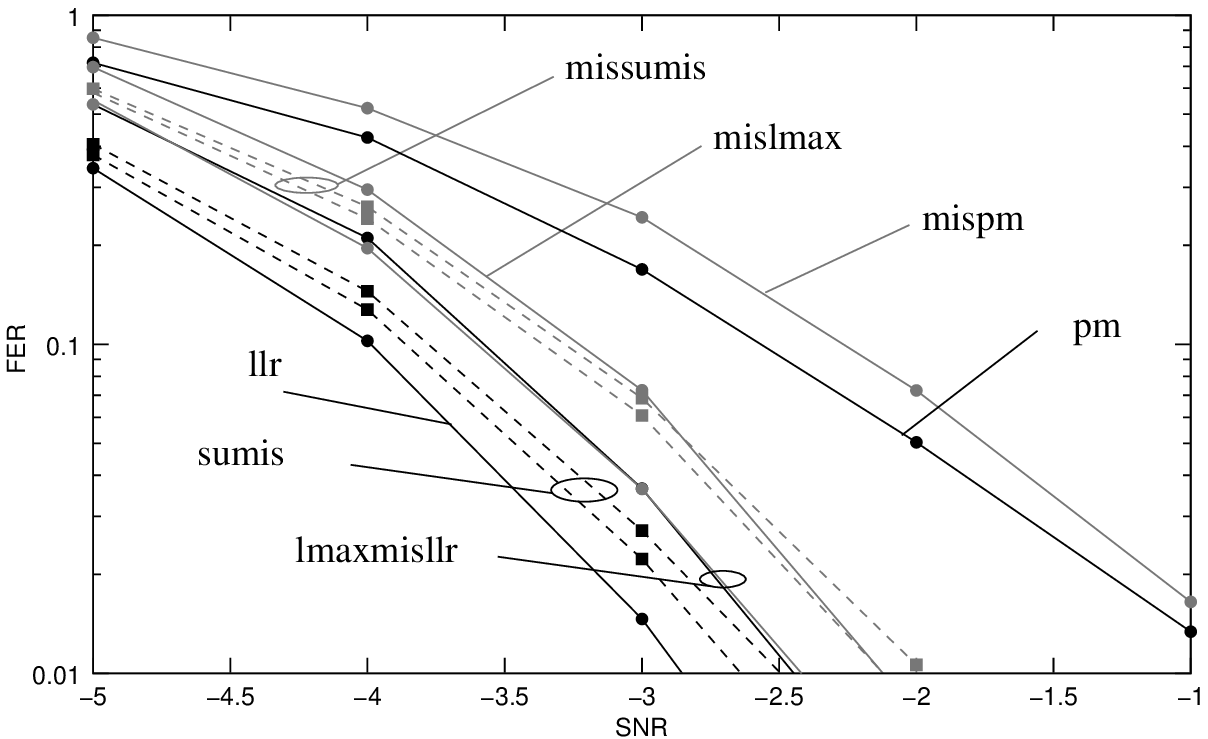}
    \caption{Same as \figref{fig:6x6} but with imperfect CSI. The
      matched SUMIS version used here is the one presented in
      \secref{sec:icsi} for constant-modulus constellations.}
    \label{fig:6x6icsiqam4}
  \end{subfigure}

  \begin{subfigure}[bt]{\columnwidth}
    \psfrag{FER}[][][\psscale][0]{FER}
    \psfrag{SNR}[][][\psscale][0]{$\Eb/\No$ [dB]}
    \psfrag{mislmax}[r][r][\psscale][0]{max-log~mismatch.}
    \psfrag{sumis}[b][b][\psscale][0]{\parbox{5cm}{\centering{SUMIS~$\ns=1,3$\\match.}}}
    \psfrag{missumis}[b][b][\psscale][0]{\parbox{5cm}{\centering{SUMIS~$\ns=1,3$\\mismatch.}}}
    \psfrag{mispm}[r][r][\psscale][0]{\parbox{2cm}{\centering{PM~$\ns=3$\\mismatch.}}}
    \includegraphics[width=\columnwidth]{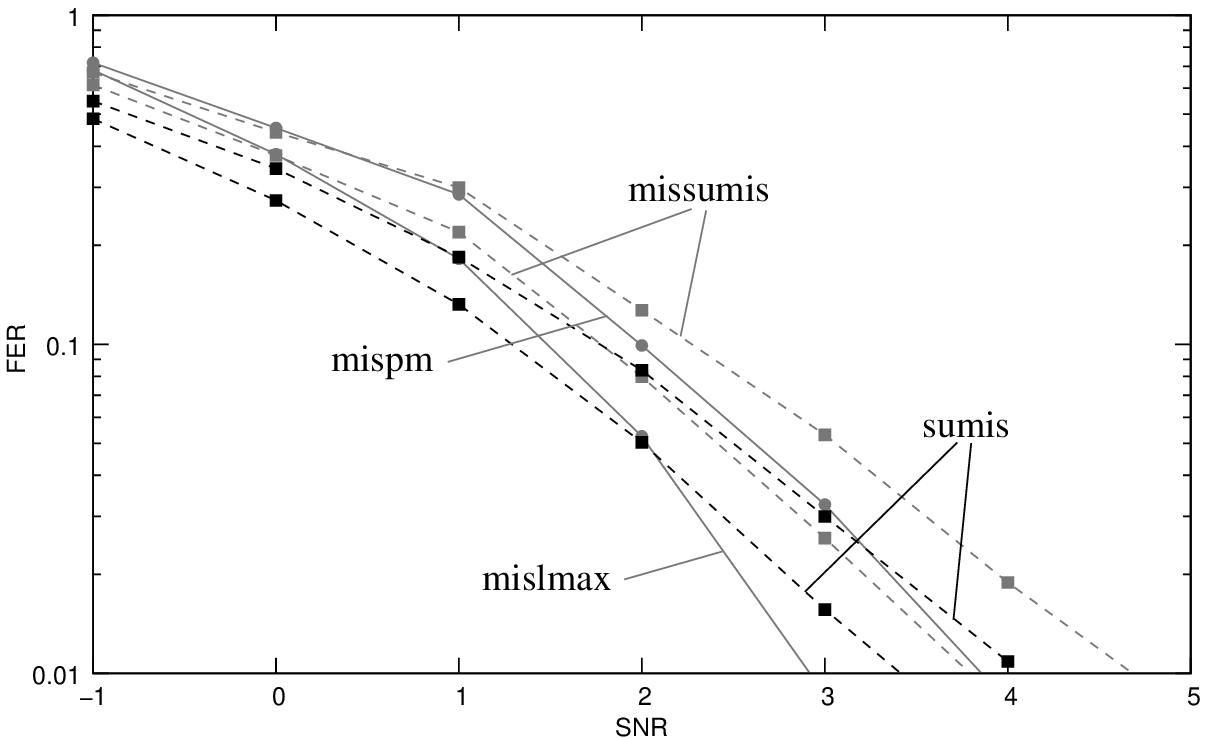}
    \caption{Same as \figref{fig:6x6qam16} but with imperfect CSI. The
      matched SUMIS version used here is presented in
      \secref{sec:icsi} for non-constant-modulus constellations.}
    \label{fig:6x6icsiqam16}
  \end{subfigure}
  \caption{An example with imperfect CSI at the receiver. The
    error-matrix-element variance $\delta^2$  is directly
    proportional to the noise variance $\No$. The matched detectors
    use $P(\bss|\bsy,\whH)$  and the mismatched use
    $P(\bss|\bsy,\bsH)\big|_{\bsH=\scrwhH}$.}
  \label{fig:6x6icsi}
\end{figure}

\begin{figure}[bt]  
  \psfrag{FER}[][][\psscale][0]{FER}
  \psfrag{SNR}[][][\psscale][0]{$\Eb/\No$ [dB]}
  \psfrag{llr}[r][r][\psscale][0]{exact LLR}
  \psfrag{lmax}[r][r][\psscale][0]{max-log}
  \psfrag{sumis}[r][r][\psscale][0]{SUMIS, $\ns=1,3$}
  \psfrag{noiter}[r][r][\psscale][0]{no iter.}
  \psfrag{iter}[r][r][\psscale][0]{3 iters.}
  \includegraphics[width=\columnwidth]{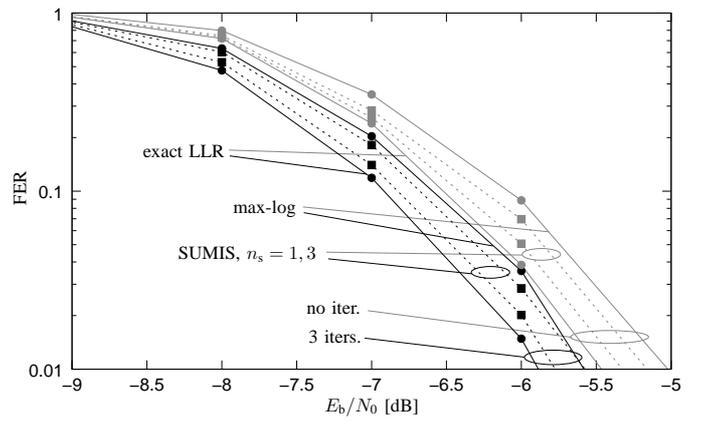}
  \caption{Detection with iterative decoding using the same setting as
    in \figref{fig:6x6}. The curves marked with gray were simulated
    with no iteration between the respective detector and the outer
    decoder. For curves marked with black, 3 iterations were used,
    which means 4 decoder runs. The extended SUMIS algorithm used here
    is presented in \secref{sec:nuniprob}.}
  \label{fig:6x6siso}
\end{figure}

In \figref{fig:6x6} we show results for a slow-fading $6\times6$ MIMO
system using $4$-QAM modulation and the LDPC code of rate $1/2$. There
is no iteration between the detector and the decoder, and the
transmitted symbols are assumed by the detector to be uniformly
distributed. This plot illustrates our principal comparison and the
rest (Figs.~\ref{fig:6x6rates}--\ref{fig:6x6siso}) illustrate extended
comparisons that deal with different scenarios of interest:
slow-/fast-fading, moderate-/large-size MIMO, low-/high-rate codes,
full/partial CSI, higher-order constellations, and
iterative/non-iterative receivers.  \figref{fig:6x6} includes all the
above mentioned detection methods whereas the remaining figures
include only those methods that show noteworthy variations from what
is already seen in \figref{fig:6x6}.

The results in \figref{fig:6x6} clearly show that the SUMIS detector
performs close to the exact LLR (optimal soft detector) performance
and it does so at a very low complexity, see also
\tabref{tab:complexity}. It outperforms the PM and max-log methods (SD
and its derivatives). Note that the complexity of SUMIS with $\ns=3$
is much lower than that of PM with $\ns=r+1=3$ even though the
partitioned problem in \eqref{eq:partmodel} is of the same size. The
reason is that the sums of the PM method consists of terms whose
exponents require the evaluation of matrix-vector multiplications of
much larger dimension than in SUMIS. Additionally, SUMIS (both the
complete algorithm and the stage-I-only variant) offers a well-defined
performance-complexity tradeoff via the choice of the parameter $\ns$.

\figref{fig:6x6rates} shows results for the same setup as in
\figref{fig:6x6} but with code rates $2/9\lesssim1/4$ and $5/6$
instead. This plot suggest that there is a larger, but still very
small, performance gap between SUMIS and exact LLR for higher coding
rates. Also, the max-log curve is much closer to the exact LLR curve
for higher rates than for lower rates. The high-rate scenario clearly
shows the importance of the model ``purification'' procedure (SUMIS
stage II) as the performance gap between SUMIS and stage-I-only SUMIS
is significant. For the low-rate scenario, the performance gap between
SUMIS and exact LLR is negligible. Similar results have been observed
for short convolution codes with a codeword length of $100$ bits, but
these plots are not included here due to space limitations. We have
also conducted similar experiments with correlated MIMO channels
(using a Toeplitz correlation structure) \cite{Vanzelst}, but these
results are omitted here due to space limitations. In these
experiments, although there was a shift in the absolute performance of
all methods, their relative performance, except for that of PM which
became much worse, did not differ significantly from what is observed
in Figs.~\ref{fig:6x6} and \ref{fig:6x6rates}. One reason why PM
performed differently may be its sensitivity to the condition number
of $\wtH$, which in the correlated MIMO setting is typically much
higher than in the uncorrelated case. This stands in contrast to SUMIS
which is much more robust in such cases.

A large MIMO system is simulated in \figref{fig:13x13}. The setup is
the same as in \figref{fig:6x6} except that here the size of the
system is $13\times13$ complex-valued (corresponding to $26\times 26$
real-valued MIMO). This figure illustrates how close both the soft
MMSE (SUMIS stage-I-only with $\ns=1$) and SUMIS are to the exact LLR
performance curve. Plots that include the exact LLR curve for large
MIMO systems are unavailable in the literature to our knowledge,
probably because of the required enormous simulation time. We used a
highly optimized version of the exact LLR detector and it took
approximately $50000$ core-hours to generate the curves in
\figref{fig:13x13}. A very interesting observation is that in
\figref{fig:13x13}, the gap between the max-log curve and the exact
LLR curve is larger than in \figref{fig:6x6}. This curve represents
the performance that the various tree-search algorithms, such as SD
with its variants and the method in \cite{Srinidhi} (which is
specifically designed for large MIMO systems), aim to achieve. As
predicted in \secref{sec:vlmimo}, we see in \figref{fig:13x13} that
the soft MMSE is much closer to the exact LLR curve than in
\figref{fig:6x6}. Also, the purification step in SUMIS yields a clear
compensation for the performance loss of soft MMSE. The performance of
SUMIS is impressive, both with stage-I-only and with stage II
included, which suggests that approximating the exact LLR expression
directly (which is the philosophy of SUMIS) is a better approach than
max-log.

Figs. \ref{fig:6x6fast} and \ref{fig:snrflops} show the results for a
fast-fading scenario. The setup is the same as in \figref{fig:6x6},
except that here each codeword spans over $40$ channel
realizations. In \figref{fig:6x6fast}, as expected, the relative
performance of simple soft MMSE (SUMIS stage-I-only for $\ns=1$) is
much better than in \figref{fig:6x6} where the channel stays constant
over a whole codeword. The reason is that the presence of
ill-conditioned channel matrices, which make linear methods such as
soft MMSE and soft ZF perform poorly, has less impact here. With
coding over many channel realizations, only a small part of each
codeword will be affected by ill-conditioned channel matrices since in
i.i.d. Rayleigh fading, they do not occur often (assuming that the
MIMO channel is not overly underdetermined). The axes in
\figref{fig:snrflops} show the number of elementary operations bundled
together (additions, subtractions, multiplications, and divisions)
versus the minimum signal-to-noise ratio required to achieve 1\%
FER. The values on the horizontal axis are calculated for all methods,
except for max-log via SD, using the expressions in
\tabref{tab:complexity}. For max-log via SD, the minimum, maximum, and
mean complexities of the $\bsy$-dependent part were calculated
empirically over many different (and random) channel and noise
realizations. More specifically, this was done for the
single-tree-search algorithm \cite{StuderBolcskei} by calculating the
number of visited nodes at different tree levels and the associated
number of elementary operations at each node (disregarding the
overhead of the book-keeping and the ``if-and-else'' statements). This
method, as presented in \cite{StuderBolcskei}, has $k$
additions/subtractions and $k$ multiplications in each node at tree
level $k$, where the root node is at level $1$. The numbers in
\figref{fig:snrflops} particularly show the advantages of SUMIS and
linear like methods such as soft MMSE.

\figref{fig:6x6qam16} reports results with $16$-QAM modulation
(instead of 4-QAM as used earlier).  Except for the modulation, all
other parameters are the same as in \figref{fig:6x6}. The results in
\figref{fig:6x6qam16} show that the performance of SUMIS is very good
and especially in relation to its complexity. For instance, for the
$\bsy$-dependent part in \tabref{tab:complexity} (having set $\ns=3$),
SUMIS, PM, and exact LLR require roughly $5.4\cdot10^3$,
$2.6\cdot10^5$, and $2\cdot10^8$ operations, respectively. This
suggests that the speedup of SUMIS is $50$ and $4\cdot10^4$ relative
to PM and to exact LLR calculation, respectively.

Yet another important scenario that often occurs in practice is
detection under imperfect CSI (ICSI). Figs.~\ref{fig:6x6icsiqam4} and
\ref{fig:6x6icsiqam16} presents performance results for this case. The
setup is the same as in \figref{fig:6x6} and \figref{fig:6x6qam16},
respectively, but here the detector is provided with knowledge of
$\whH$ instead of $\bsH$. The error-matrix-element variance $\delta^2$
is proportional to the noise variance $\No$, i.e.,
$\delta^2=\alpha\No$ where $\alpha$ is a constant. For SUMIS, we
considered both the intelligent way of handling ICSI (using
$P(\bss|\bsy,\whH)$) and the crude way using mismatched detection
(inserting $\whH$ into $P(\bss|\bsy,\bsH)$), see
\secref{sec:icsi}. For the other detectors, we considered only the
mismatched detector since versions of those algorithms that perform
intelligent detection using $P(\bss|\bsy,\whH)$ and higher-order
constellations do not seem to be available.  Clearly, intelligently
handling ICSI yields better performance than performing mismatched
detection.  The results in \figref{fig:6x6icsiqam4} resemble those in
\figref{fig:6x6} with only a minor shift in signal-to-noise
ratio. This comes as no surprise as the effective channel model in
\eqref{eq:icsimodel} for BPSK per real dimension (as in
\figref{fig:6x6icsiqam4}) is equivalent to \eqref{eq:model} (as in
\figref{fig:6x6}) up to a scaling of the noise variance.

Finally, \figref{fig:6x6siso} shows results with iterative decoding
where the detector and the decoder interchange information, hence
exploiting the soft-input capability of SUMIS. The iterative decoding
setup is that of \cite[fig. 1]{WangPoor}, and the simulation model is
the same as used in \figref{fig:6x6}. SUMIS, as presented in
\secref{sec:nuniprob}, shows strikingly good performance at very low
complexity.

\section{Conclusions}

We have proposed a novel soft-input soft-output MIMO detection method,
SUMIS, that outperforms today's state-of-the-art detectors (such as
PM, FCSD, SD and its derivatives), runs at fixed-complexity, provides
a clear and well-defined tradeoff between computational complexity and
detection performance, and is highly parallelizable. The ideas behind
SUMIS are fundamentally simple and allow for very simple algorithmic
implementations. The proposed method has a complexity comparable to
that of linear detectors. We have conducted a thorough numerical
performance evaluation of our proposed method and compared to
state-of-the-art methods. The results indicate that in many cases
SUMIS (for low $\ns$) outperforms the max-log method and therefore
inherently all other methods that approximate max-log, such as SD and
its derivatives. This performance is achieved with a complexity that
is much smaller than that of competing methods, see
\tabref{tab:complexity} and \figref{fig:snrflops}, and in particular
smaller than the initialization step of SD alone. In terms of hardware
implementation, SUMIS has remarkable advantages over tree-search
(e.g., SD) algorithms that require comparisons and branchings
(if-then-else statements).

More fundamentally, the results indicate that approximating the exact
LLR expression directly (which is the basic philosophy of SUMIS) is a
better approach than max-log followed by hard-decisions. This way of
thinking, similar to \cite{LarssonJalden}, represents a profound shift
in the way the problem is tackled and opens the door for further new
approaches to the detection problem. The increase in performance that
this philosophy offers is especially pronounced in larger MIMO systems
where the performance gap between the max-log approximation and the
exact LLR seems to increase.

\appendix

\subsection{Optimized SUMIS and Soft MMSE and Their Complexity}
\label{app:cmplx}

\begin{figure*}[hbt]
\hrule
\begin{equation}
\label{eq:completesq}
\begin{split}
\norm{\bar{\bsy}-\wbH\bar{\bss}}^2_\bsQ%
&=(\bar{\bsy}-\wbH\bar{\bss})^T\bsQ^{-1}(\bar{\bsy}-\wbH\bar{\bss})%
=\bar{\bsy}^T\bsQ^{-1}\bar{\bsy}-2\bar{\bsy}^T\bsQ^{-1}\wbH\bar{\bss}+\bar{\bss}^T\wbH^T\bsQ^{-1}\wbH\bar{\bss}\\
&=((\wbH^T\bsQ^{-1}\wbH)^{-1}\wbH^T\bsQ^{-1}\bar{\bsy}-\bar{\bss})^T\wbH^T\bsQ^{-1}\wbH((\wbH^T\bsQ^{-1}\wbH)^{-1}\wbH^T\bsQ^{-1}\bar{\bsy}-\bar{\bss})\\
&\quad\quad+\bar{\bsy}^T\bsQ^{-1}\bar{\bsy}-\bar{\bsy}^T\bsQ^{-1}\wbH(\wbH^T\bsQ^{-1}\wbH)^{-1}\wbH^T\bsQ^{-1}\bar{\bsy}
\end{split}
\end{equation}
\hrule

\def\NoTwo{\text{\makebox[3mm]{\scriptsize$\frac{\No}{2}$}}}
\begin{subequations}
\label{eq:matinvident}
\begin{equation}
\label{eq:matinvhq}
\begin{split}
\wbH^T\!\bsQ^{-1}&=\wbH^T\!(\NoTwo\bsI+\wtH\wtPsi\wtH^T\!)^{-1}%
=\wbPsi^{-1}\wbPsi\wbH^T(\NoTwo\bsI+\wtH\wtPsi\wtH^T)^{-1}\\
&=\wbPsi^{-1}\!\!\underbrace{\wbPsi\wbH^T\!(\NoTwo\bsI+\bsH\bsPsi\bsH^T\!\!-\wbH\wbPsi\wbH^T)^{-1}}_{\text{apply
    matrix inversion lemma}}\\
&=\wbPsi^{-1}\big(\wbPsi^{-1}\!\!-\wbH^T\!(\NoTwo\bsI+\bsH\bsPsi\bsH^T)^{-1}\wbH\big)^{-1}%
\wbH^T\!(\NoTwo\bsI+\bsH\bsPsi\bsH^T)^{-1}\\
&=\big(\bsI-\wbH^T\!(\NoTwo\bsI+\bsH\bsPsi\bsH^T)^{-1}\wbH\wbPsi\big)^{-1}%
\wbH^T\!(\NoTwo\bsI+\bsH\bsPsi\bsH^T)^{-1}
\end{split}
\end{equation}

\begin{equation}
\label{eq:matinvhqh}
\begin{split}
\wbH^T\!\bsQ^{-1}\wbH&\overset{\eqref{eq:matinvhq}}{=}%
\big(\bsI-\wbH^T\!(\NoTwo\bsI+\bsH\bsPsi\bsH^T)^{-1}\wbH\wbPsi\big)^{-1}%
\wbH^T\!(\NoTwo\bsI+\bsH\bsPsi\bsH^T)^{-1}\wbH=\\
&\Big(\big(\wbH^T\!(\NoTwo\bsI+\bsH\bsPsi\bsH^T)^{-1}\wbH\big)^{-1}-\wbPsi\Big)^{-1}
\end{split}
\end{equation}
\end{subequations}
\hrule\vspace{1em}

\begin{subequations}
\label{eq:matinvh}
\begin{equation}
\label{eq:matinvhqhihq}
(\wbH^T\!\bsQ^{-1}\wbH)^{-1}\wbH^T\!\bsQ^{-1}\overset{\eqref{eq:matinvident}}{=}%
\big(\wbH^T\!(\NoTwo\bsI+\bsH\bsPsi\bsH^T)^{-1}\wbH\big)^{-1}\wbH^T\!(\NoTwo\bsI+\bsH\bsPsi\bsH^T)^{-1}
\end{equation}

\begin{equation}
\label{eq:phiplushdh}
\begin{split}
\wbH^T\!(\NoTwo\bsI+\bsH\bsPsi\bsH^T)^{-1}&=\wbP^T\!\!\bsH^T\!(\NoTwo\bsI+\bsH\bsPsi\bsH^T)^{-1}%
=\wbP^T\!\!\bsPsi^{-1}\!\underbrace{\bsPsi\bsH^T\!(\NoTwo\bsI+\bsH\bsPsi\bsH^T)^{-1}}_{\text{apply matrix inversion lemma}}\\
&=\wbP^T\!\!\bsPsi^{-1}\!(\,\text{\makebox[3mm]{\scriptsize$\frac{\No}{2}$}}\bsPsi^{-1}\!\!+\bsH^T\!\!\bsH)^{-1}\!\bsH^T
\end{split}
\end{equation}
\end{subequations}
\hrule
\end{figure*}

We first identify the main complexity bottlenecks of SUMIS step by
step and keep track of the largest order of magnitude terms. The focus
will be on \algref{alg:sumis} in \secref{sec:sumis}, its optimization
and the number of operations required for its execution. Note that the
techniques presented in what follows can also be used for an optimized
implementation of the soft MMSE method \cite{StuderFateh}. The
complexity will be measured in terms of elementary operations bundled
together: additions, subtractions, multiplications, and divisions. The
complexity count is divided into two parts: a received data
independent ($\bsy$-independent) processing part and an
$\bsy$-dependent processing part. We will also assume that
$\ns\ll\Nt\leq\Nr$, which is the case of most practical interest. The
assumption on $\Nt\leq\Nr$ is required only for the presented
optimized SUMIS version, due to the requirement for various inverses
to exist, but analogous complexity reductions can be made for
$\Nt\geq\Nr$ and are excluded due to space limitations.

\subsubsection{$\bsy$-independent processing}

We start with the choice of permutations in
\secref{ssec:sumisperms}. The SUMIS algorithm uses $\Nt$ different
permutations that are decided based on $\bsH^T\bsH$. This procedure
evaluates $\bsH^T\bsH$ requiring $\Nt^2\Nr$ operations. There is also
a small search involved that requires $\Nt(\ns-1)(\Nt-\ns)$
comparisons and this we neglect.

Next, by simple matrix manipulations, one can pre-process and simplify
the computation of \eqref{eq:skprob}, in the $\bsy$-dependent part of
the algorithm, consisting of a sum of terms in \eqref{eq:prbybarsbar}
over all $\bar{\bss}$. Consider again \eqref{eq:prbybarsbar},
\begin{equation}
p(\bar{\bsy}|\bar{\bss})=\frac{1}{\sqrt{(2\pi)^\Nr|\bsQ|}}%
\exp{-\frac{1}{2}\norm{\bar{\bsy}-\wbH\bar{\bss}}^2_{\bsQ}},
\end{equation}
which includes matrix-vector multiplications of dimension $\Nr$. We
can rewrite the exponent as in \eqref{eq:completesq} where the terms
on the last line in \eqref{eq:completesq} do not depend on
$\bar{\bss}$ and will not affect the final result in
\eqref{eq:skprob}. So, from \eqref{eq:completesq}, we see that if
$\wbH^T\bsQ^{-1}\wbH$ and $(\wbH^T\bsQ^{-1}\wbH)^{-1}\wbH^T\bsQ^{-1}$
are precomputed, the matrix-vector multiplications in
\eqref{eq:prbybarsbar} in the $\bsy$-dependent part will be of
dimension $\ns\ll\Nr$, which is evidently desirable.

We need to evaluate these matrices once for each partitioning (there are
$\Nt$ of them). This can be done simultaneously in one step. For this purpose, we
derive the identities in \eqref{eq:matinvident} and \eqref{eq:matinvh}
where we have defined $\bsPsi$ and $\wbPsi$ to be diagonal matrices
such that $\bsH\bsPsi\bsH^T=\wbH\wbPsi\wbH^T+\wtH\wtPsi\wtH^T$, and
$\wbP\in\{0,1\}^{\Nt\times\ns}$ to be a matrix that has precisely
$\ns$ ones such that $\wbH=\bsH\wbP$ (a column picking matrix). Recall
that $\wtPsi$ is the covariance matrix of $\tilde{\bss}$. For $\ns=1$,
the identity \eqref{eq:matinvhq} (also mentioned in
\cite{StuderFateh,Studer}) is well known from the equivalence showed in
\cite[exerc. 8.18]{TseViswanath} of the MMSE filter
\cite[sec. 3.2.1]{KailathSayed}. Equation \eqref{eq:phiplushdh} was
also derived in \cite{StuderFateh} and \cite{Studer} though the
derivations there contain a minor error. Specifically, the assumption
on $\bsU\bsU^T=\bsI$ in the singular value decomposition of
$\bsH=\bsU\bsSigma\bsV$ in \cite[app. A.2.2]{Studer} is not valid for
$\Nr>\Nt$.

Now, since we have established \eqref{eq:matinvhqh} and
\eqref{eq:phiplushdh}, we can immediately write
\begin{equation}
\label{eq:hqinvh}
\wbH^T\!\!\inv{\bsQ}\wbH=%
\Big(\big(\wbP^T\!\!\inv{\bsPsi}\inv{(\,\text{\makebox[3mm]{\scriptsize$\frac{\No}{2}$}}\inv{\bsPsi}+%
\bsH^T\!\!\bsH)}\!\bsH^T\!\!\bsH\wbP\big)^{-1}\!\!\!-\wbPsi\Big)^{-1},
\end{equation}
where the innermost inverse is of dimension $\Nt$ and the two
outermost inversions are of dimension $\ns$. Focusing on the innermost
inverse, it has been observed in \cite{StuderFateh} that the matrix
$(\,\text{\makebox[3mm]{\scriptsize$\frac{\No}{2}$}}\inv{\bsPsi}+\bsH^T\!\!\bsH)$
can be numerically unstable to invert. The reason is that some
(diagonal) values in $\bsPsi$ can be very small. This was addressed in
\cite{StuderFateh} by writing $\inv{\bsPsi}%
\inv{(\,\text{\makebox[3mm]{\scriptsize$\frac{\No}{2}$}}\inv{\bsPsi}+\bsH^T\!\!\bsH)}=%
\inv{(\,\text{\makebox[3mm]{\scriptsize$\frac{\No}{2}$}}\bsI+%
  \bsH^T\!\!\bsH\bsPsi)}$, which is a more stable inverse but due to
the lost symmetry property requires much more operations
\cite{GolubVanloan}. We want to facilitate the use of efficient
algorithms available for inversion of symmetric matrices
\cite{GolubVanloan} but without having to deal with unstable
inversions. Therefore, we instead write
$\inv{\bsPsi}\inv{(\,\text{\makebox[3mm]{\scriptsize$\frac{\No}{2}$}}\inv{\bsPsi}+\bsH^T\!\!\bsH)}=%
\bsPsi^{-\frac{1}{2}}\!\inv{(\,\text{\makebox[3mm]{\scriptsize$\frac{\No}{2}$}}\bsI+%
  \bsPsi^{\frac{1}{2}}\!\bsH^T\!\!\bsH\bsPsi^{\frac{1}{2}})}\bsPsi^{\frac{1}{2}}$
where $(\,\text{\makebox[3mm]{\scriptsize$\frac{\No}{2}$}}\bsI+%
\bsPsi^{\frac{1}{2}}\!\!\bsH^T\!\!\bsH\bsPsi^{\frac{1}{2}})$ is
symmetric and stable to invert. Note that the computation of
$\bsPsi^{\frac{1}{2}}$ and $\bsPsi^{-\frac{1}{2}}$ is simple, even
though it necessitates square root evaluations, since $\bsPsi$ is
diagonal with positive values.

There are several different approaches to inverting a positive
definite matrix. Some are more numerically stable than others and some require
less operations than others. One very fast and stable approach is
through the LDL-decomposition \cite[p.139]{GolubVanloan}, i.e.,
$(\,\text{\makebox[3mm]{\scriptsize$\frac{\No}{2}$}}\bsI+%
\bsPsi^{\frac{1}{2}}\!\bsH^T\!\!\bsH\bsPsi^{\frac{1}{2}})=\bsL\bsD\bsL^T$,
where $\bsL$ is a lower-triangular matrix with ones on its diagonal
and $\bsD$ is a diagonal matrix with positive diagonal elements. The
LDL-decomposition itself requires $\Nt^3/3$ operations
\cite[p.139]{GolubVanloan}, and so does the inversion of $\bsL$ and
$\bsD$ together. Hence, \eqref{eq:hqinvh} becomes
\begin{equation}
\wbH^T\!\!\inv{\bsQ}\wbH=%
\Big(\big(\wbP^T\!\!\bsPsi^{-\frac{1}{2}}\!\bsL^{-T}\!\!\inv{\bsD}%
\inv{\bsL}\bsPsi^{\frac{1}{2}}\!(\bsH^T\!\!\bsH)\wbP\big)^{-1}\!\!\!-\wbPsi\Big)^{-1},
\end{equation}
for which the number of operations for all partitionings can be
summarized:
\begin{itemize}
  \item{LDL-decomposition ($\Nt^3/3$),}
  \item{$\inv{\bsL}$ ($\Nt^3/3$),}
  \item{$\bsPsi^{-\frac{1}{2}}\!\bsL^{-T}\!\inv{\bsD}\inv{\bsL}\bsPsi^{\frac{1}{2}}$
    ($\Nt^3/3$),}
  \item{$\big(\wbP^T\!\bsPsi^{-\frac{1}{2}}\!\!\bsL^{-T}\!\!\inv{\bsD}\inv{\bsL}%
    \bsPsi^{\frac{1}{2}}\!(\bsH^T\!\!\bsH)\wbP\big)$ $2\ns^2\Nt^2$.}
\end{itemize}
The remaining evaluations consist of inverses of matrices of very
small dimension $\ns$ for which there exist closed form formulas that
require a negligible number of operations. Thus, the total number of
operations required to compute $\wbH^T\!\!\inv{\bsQ}\wbH$ explicitly
for all partitionings is $\Nt^3+2\ns^2\Nt^2$.

\subsubsection{$\bsy$-dependent processing}
\label{app:cmplx:ydep}

We need to compute, for all partitionings,
\begin{equation}
\label{eq:hqinvhy}
\begin{split}
\inv{(\wbH^T\!\!\inv{\bsQ}\wbH)}\!\!\wbH^T\!\!\inv{\bsQ}\bsy\overset{\eqref{eq:matinvh}}{=}%
\big(&\wbP^T\!\!\bsPsi^{-\frac{1}{2}}\!\bsL^{-T}\!\!\inv{\bsD}%
\inv{\bsL}\bsPsi^{\frac{1}{2}}\!\bsH^T\!\!\bsH\wbP\big)^{-1}\\
&\times\wbP^T\!\!\bsPsi^{-\frac{1}{2}}\!\bsL^{-T}\!\!\inv{\bsD}%
\inv{\bsL}\bsPsi^{\frac{1}{2}}\!\bsH^T\!\!\bsy,
\end{split}
\end{equation}
where only $\wbP^T\!\!\bsPsi^{-\frac{1}{2}}\!\bsL^{-T}\!\!\inv{\bsD}%
\inv{\bsL}\bsPsi^{\frac{1}{2}}\!\bsH^T\!\!\bsy$ needs to
be evaluated since the leftmost (inverse) matrix in
\eqref{eq:hqinvhy} is of dimension $\ns$ and has already been computed
in the $\bsy$-independent part. Note that the computation of
$\bsH^T\!\!\bsy$ and subsequently
$\bsPsi^{-\frac{1}{2}}\!\bsL^{-T}\!\!\inv{\bsD}%
\inv{\bsL}\bsPsi^{\frac{1}{2}}\!\bsH^T\!\!\bsy$ requires $2\Nr\Nt$
and $2\Nt^2$ operations, respectively. Hence, to compute
$(\wbH^T\!\!\inv{\bsQ}\wbH)^{-1}\!\!\wbH^T\inv{\bsQ}\bsy$ for all
partitionings requires $2\Nr\Nt+2\Nt^2$ operations.

To compute \eqref{eq:skprob}, for each $s_k$, requires $\ns^22^\ns$
operations since the exponents in \eqref{eq:prbybarsbar} consist of
matrix-vector multiplications of dimension $\ns$. This, we can safely
neglect when $2^\ns\ll\Nt^2$ (which is typically the case). If
higher-order constellations with cardinality much higher than 2 are
used, one can keep the cardinality small and fixed by disregarding
constellation points outside an appropriately chosen ellipse centered
at the mean.

The remaining bottleneck is the computation of the updated covariance
matrix $\wbH^T\!\!\inv{\bsQ'}\wbH$ and the update of $\bsy$ to $\bsy'$
in \eqref{eq:suppmodel}. The number of operations required for
$\wbH^T\!\!\inv{\bsQ'}\wbH$ is $\Nt^3+2\ns^2\Nt^2$, analogously to the
computation of $\wbH^T\!\!\inv{\bsQ}\wbH$. For the update in
\eqref{eq:suppmodel}, we have that
$\bsy'=\bsy-\wtH\Exp{\tilde{\bss}|\bsy}=%
\bsy-\bsH\Exp{\bss|\bsy}+\wbH\Exp{\bar{\bss}|\bsy}$, which after the
transformation in \eqref{eq:hqinvhy} using the updated matrix $\bsQ'$
instead of $\bsQ$ becomes
\begin{align}
\label{eq:hqinvhyprim}
\hspace{2mm}&\hspace{-2mm}(\wbH^T\!\!\inv{\bsQ'}\wbH)^{-1}\!\!\wbH^T\!\!\inv{\bsQ'}\bsy'\notag\\
&=\inv{(\wbH^T\!\!\inv{\bsQ'}\wbH)}\!\!\wbH^T\!\!\inv{\bsQ'}(\bsy-\bsH\Exp{\bss|\bsy})+\Exp{\bar{\bss}|\bsy}\notag\\
&=\Exp{\bar{\bss}|\bsy}+\Big(\big(\wbP^T\!\!\inv{\bsPhi}%
(\,\text{\makebox[3mm]{\scriptsize$\frac{\No}{2}$}}\inv{\bsPhi}+%
\bsH^T\!\!\bsH)^{-1}\!\bsH^T\!\!\bsH\wbP\big)^{-1}\notag\\
&\quad\quad\times\!\wbP^T\!\!\inv{\bsPhi}%
(\,\text{\makebox[3mm]{\scriptsize$\frac{\No}{2}$}}\inv{\bsPhi}+%
\bsH^T\!\!\bsH)^{-1}\!\big(\bsH^T\!\!\bsy-\bsH^T\!\!\bsH\Exp{\bss|\bsy}\!\big)\Big).
\end{align}
The relation between the matrices $\bsPhi$ and $\tilde{\bsPhi}$ is
analogous to the relation between $\bsPsi$ and $\tilde{\bsPsi}$. From
the discussion after \eqref{eq:hqinvhy}, we can conclude that
\eqref{eq:hqinvhyprim} requires $4\Nt^2$ operations for all
partitionings.  Lastly, the LLR computation of each bit requires
$\ns^22^\ns$ operations, which we can safely neglect when $2^\ns\ll\Nt^2$.

\subsubsection{Summary}

Under the assumption that $\Nr\geq\Nt$ and that no a priori knowledge
of $\bss$ is available, the $\bsy$-independent part of the algorithm
requires roughly $\Nr\Nt^2+\Nt^3+2\ns^2\Nt^2$ operations, which is a
similar number of operations as required by the soft MMSE
algorithm. As for the $\bsy$-dependent part, the number of operations
required is roughly $\Nt^3+2\Nr\Nt+(2\ns^2+6)\Nt^2$. Thus, the total
number of operations required by the SUMIS detector to evaluate all
LLRs associated with one received vector $\bsy$ is
\begin{equation}
\Nr\Nt^2+2\Nt^3+(4\ns^2+6)\Nt^2+2\Nr\Nt.
\end{equation}

The processing in SUMIS that is performed per bit can be done in
parallel. The processing (per channel matrix) that involves matrix
decompositions and inversions is not as simple to parallelize. More
specifically, the LDL (or equivalently the Cholesky) decomposition has
an inherent sequential structure that cannot be fully
parallelized. Such sequential structures are present in most matrix
algebraic operations (e.g., inversions) that are commonly used in
detection algorithms. While those operations cannot be fully
parallelized, they can be highly parallelized, see \cite{Becker} and
the references therein.

\bibliographystyle{IEEEbib}
\bibliography{refs}

%% \newpage~
%% \def\biofigsize{0.35\columnwidth}
%% \small
%% \input{bios/bio-mirsad}

%% \input{bios/bio-erik}

\end{document}